\documentclass[10pt,twocolumn,letterpaper]{article}

\usepackage{cvpr}
\usepackage{times}
\usepackage{epsfig}
\usepackage{graphicx}
\usepackage{amsmath}
\usepackage{amssymb}
\usepackage{subfig}
\usepackage{float}
\usepackage{math}

\usepackage{xcolor}
\usepackage{multirow}
\usepackage{booktabs}
\usepackage{tikz}
\usepackage{array}
\usepackage{tabularx}
\usepackage{enumitem}
\usepackage{cuted}

\usepackage{csquotes}

\def\xgen{\hat{\xvect}^\tau}

\def\xgen{\hat{\xvect}'}
\def\netGen{{G}}
\def\netDet{{\Delta}}

\def\eg{\textit{e.g.}\xspace}
\def\ie{\textit{i.e.}\xspace}

\def\etal{\textit{et al.}\xspace}

\usepackage[breaklinks=true,bookmarks=false]{hyperref}
\hypersetup{
    colorlinks=true,
    filecolor=magenta,      
    urlcolor=magenta,
}
\cvprfinalcopy 

\def\cvprPaperID{4908} 

\begin{document}


\title{
Animating Arbitrary Objects 
via Deep Motion Transfer
}

\author{Aliaksandr Siarohin$^1$, St{\'e}phane Lathuili{\`e}re$^1$, Sergey Tulyakov$^2$, Elisa Ricci$^{1,3}$ and Nicu Sebe$^{1,4}$\\
$^1$DISI, University of Trento, Italy, $^2$ Snap Inc., Santa Monica, CA, \\
$^3$  Fondazione Bruno Kessler (FBK), Trento, Italy,\\
$^4$Huawei Technologies Ireland, Dublin, Ireland\\
{\tt\small \{aliaksandr.siarohin,stephane.lathuilire,e.ricci,niculae.sebe\}@unitn.it}, {\tt\small stulyakov@snap.com}}
\maketitle


\begin{abstract}
\vspace{-0.2cm}
   This paper introduces a novel deep learning framework for image animation. Given an input image with a target object and a driving video sequence depicting a moving object, our framework generates a video in which the target object is \textit{animated} according to the driving sequence. This is achieved through a deep architecture that decouples appearance and motion information. Our framework consists of three main modules: (i) a Keypoint Detector unsupervisely trained to extract object keypoints, (ii) a Dense Motion prediction network for generating dense heatmaps from sparse keypoints, in order to better encode motion information and (iii) a Motion Transfer Network, which uses the motion heatmaps and appearance information extracted from the input image to synthesize the output frames.
   We demonstrate the effectiveness of our method on several benchmark datasets, spanning a wide variety of object appearances, and show that our approach outperforms state-of-the-art image animation and video generation methods. Our source code is publicly  available~\footnote{ \href{https://github.com/AliaksandrSiarohin/monkey-net}{https://github.com/AliaksandrSiarohin/monkey-net}}.
\end{abstract}


\vspace{-.8cm}
\section{Introduction}
\begin{figure}[t]
\begin{center}
\includegraphics[width=0.9\columnwidth]{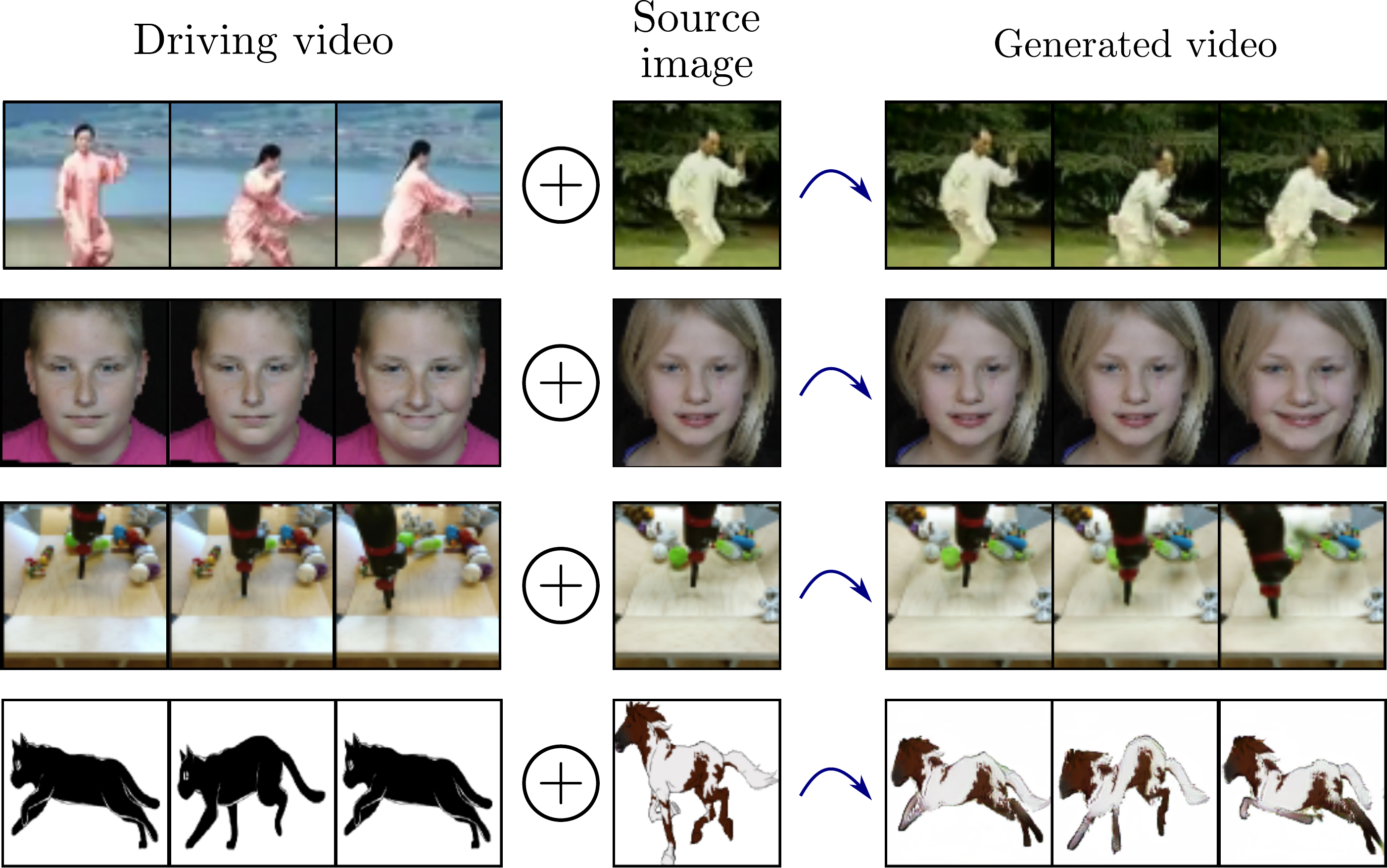}
\caption{Our deep motion transfer approach can animate arbitrary objects following the motion of the driving video.}
\label{fig:teaser}
\vspace{-0.9cm}
\end{center}
\end{figure}

\vspace{-0.2cm}
\label{Introduction}
This paper introduces a framework for motion-driven image animation to automatically generate videos by combining the appearance information derived from a \textit{source image} (\eg depicting the face or the body silhouette of a certain person) with motion patterns extracted from a \textit{driving video} (\eg encoding the facial expressions or the body movements of another person). Several examples are given in Fig.~\ref{fig:teaser}.
Generating high-quality videos from static images is challenging, as it requires learning an appropriate representation of an object, such as a 3D model of a face or a human body. This task also requires accurately extracting the motion patterns from the driving video and mapping them on the object representation. Most approaches are object-specific, using techniques from computer graphics \cite{cao2014displaced,thies2016face2face}. These methods also use an explicit object representation, such as a 3D morphable model \cite{blanz1999morphable}, to facilitate animation, and therefore only consider faces. 

Over the past few years, researchers have developed approaches for automatic synthesis and enhancement of visual data. Several methods derived from Generative Adversarial Networks (GAN) \cite{goodfellow2014generative} and Variational Autoencoders (VAE) \cite{kingma2013auto} have been proposed to generate images and videos \cite{pix2pix2016,siarohin2018whitening,saito2017temporal,tulyakov2017mocogan,roy2018semantic,tang2019dual,tang2019selection,Siarohin_2018_ACCV}. These approaches use additional information such as conditioning labels (\eg indicating a facial expression, a body pose) \cite{wang2018every,siarohin2018deformable,zhenglin2019guided,hao2018Geaturegan}. More specifically, they are purely data-driven, leveraging a large collection of training data to learn a latent representation of the visual inputs for synthesis. 
Noting the significant progress of these techniques, recent research studies have started exploring the use of deep generative models for image animation and video retargeting \cite{wiles2018x2face,chan2018everybody,bansal2018recycle,wang2018video, balakrishnansynthesizing}. 
These works demonstrate that deep models can effectively transfer motion patterns between human subjects in videos \cite{bansal2018recycle}, or transfer a facial expression from one person to another \cite{wiles2018x2face}. However, these approaches have limitations: for example, they rely on pre-trained models for extracting object representations that require costly ground-truth data annotations \cite{chan2018everybody, wang2018video, balakrishnansynthesizing}. Furthermore, these works do not address the problem of animating arbitrary objects: instead, considering a single object category \cite{wiles2018x2face} or learning to translate videos from one specific domain 
to another \cite{bansal2018recycle,joo2018generating}.

This paper addresses some of these limitations by introducing a novel deep learning framework for animating a static image using a driving video. Inspired by \cite{wiles2018x2face}, we propose learning a latent representation of an object category in a self-supervised way, leveraging a large collection of video sequences. There are two key distinctions between our work and \cite{wiles2018x2face}. Firstly, our approach is not designed for specific object category, but rather is effective in animating arbitrary objects. Secondly, we introduce a novel strategy to model and transfer motion information, using a set of sparse motion-specific keypoints that were learned in an unsupervised way to describe relative pixel movements. 
Our intuition is that only relevant motion patterns (derived from the driving video) must be transferred for object animation, while other information should not be used. 
We call the proposed deep framework Monkey-Net, as it enables motion transfer by considering MOviNg KEYpoints.

We demonstrate the effectiveness of our framework by conducting an extensive experimental evaluation on three publicly available datasets, previously used for video generation: the Tai-Chi \cite{tulyakov2017mocogan}, the BAIR robot pushing \cite{ebert2017self} and the UvA-NEMO Smile \cite{dibekliouglu2012you} datasets. As shown in our experiments, our image animation method produces high quality videos for a wide range of objects. 
Furthermore, our quantitative results clearly show that our approach outperforms state-of-the-art methods for image-to-video translation tasks.




\vspace{-0.2cm}
\section{Related work}
\vspace{-0.2cm}
\label{Related}

\textbf{Deep Video Generation.}
Early deep learning-based approaches for video generation proposed synthesizing videos by using spatio-temporal networks. Vondrick \textit{et al.}~\cite{vondrick2016generating} introduced VGAN, a 3D convolutional GAN which simultaneously generates all the frames of the target video.
Similarly, Saito \textit{et al.} \cite{saito2017temporal} proposed TGAN, 
a GAN-based model which is able to generate multiple frames at the same time. However, the visual quality of these methods outputs is typically poor.  

More recent video generation approaches used recurrent neural networks within an adversarial training framework. For instance, Wang \textit{et al.} \cite{wang2018every} introduced a Conditional MultiMode Network (CMM-Net), a deep architecture which adopts a conditional Long-Short Term Memory (LSTM) network and a VAE to generate face videos. Tulyakov \textit{et al.} \cite{tulyakov2017mocogan} proposed MoCoGAN, a deep architecture based on a recurrent neural network trained with an adversarial learning scheme. 
These approaches can take conditional information as input that comprises categorical labels or static images and, as a result, produces high quality video frames of desired actions. 

Video generation is closely related to the future frame prediction problem addressed in \cite{srivastava2015unsupervised,oh2015action,finn2016unsupervised,van2017transformation,zhao2018learning}. Given a video sequence, these methods aim to synthesize a sequence of images which represents a coherent continuation of the given video. Earlier methods \cite{srivastava2015unsupervised,oh2015action,kalchbrenner2016video} attempted to directly predict the raw pixel values in future frames. Other approaches \cite{finn2016unsupervised,van2017transformation,babaeizadeh2017stochastic} proposed learning the transformations which map the pixels in the given frames to the future frames. Recently, Villegas~\etal~\cite{villegas2017learning} introduced a hierarchical video prediction model consisting of two stages: it first predicts the motion of a set of landmarks using an LSTM, then generates images from the landmarks.

Our approach is closely related to these previous works since we also aim to generate video sequences by using a deep learning architecture. However, we tackle a more challenging task: image animation requires decoupling and modeling motion and content information, as well as a recombining them.


\textbf{Object Animation.} Over the years, the problems of image animation and video re-targeting have attracted attention from many researchers in the fields of computer vision, computer graphics and multimedia. Traditional approaches \cite{cao2014displaced,thies2016face2face} are designed for specific domains, as they operate only on faces, human silhouettes, \etc. In this case, an explicit representation of the object of interest is required to generate an animated face corresponding to a certain person's appearance, but with the facial expressions of another.
For instance, 3D morphable models \cite{blanz1999morphable} have been traditionally used for face animation \cite{zollhofer2018state}. While especially accurate, these methods are highly domain-specific and their performance drastically degrades in challenging situations, such as in the presence of occlusions.

Image animation from a driving video can be interpreted as the problem of transferring motion information from one domain to another. 
Bansal \etal~\cite{bansal2018recycle} proposed Recycle-GAN, an approach which extends conditional GAN by incorporating spatio-temporal cues in order to generate a video in one domain given a video in another domain. However, their approach only learns the association between two specific domains, while we want to animate an image depicting one object without knowing at training time which object will be used in the driving video. Similarly, Chan \etal \cite{chan2018everybody} addressed the problem of motion transfer, casting it within a per-frame image-to-image translation framework. They also proposed incorporating spatio-temporal constraints. The importance of considering temporal dynamics for video synthesis was also demonstrated in \cite{wang2018video}.
Wiles \etal \cite{wiles2018x2face} introduced X2Face, a deep architecture which, given an input image of a face, modifies it according to the motion patterns derived from another face or another modality, such as audio. They demonstrated that a purely data-driven deep learning-based approach is effective in animating still images of faces without demanding explicit 3D representation. In this work, we design a self-supervised deep network for animating static images, which is effective for generating arbitrary objects. 



\vspace{-0.2cm}
\begin{figure*}[t]\centering
\includegraphics[width=0.90\linewidth]{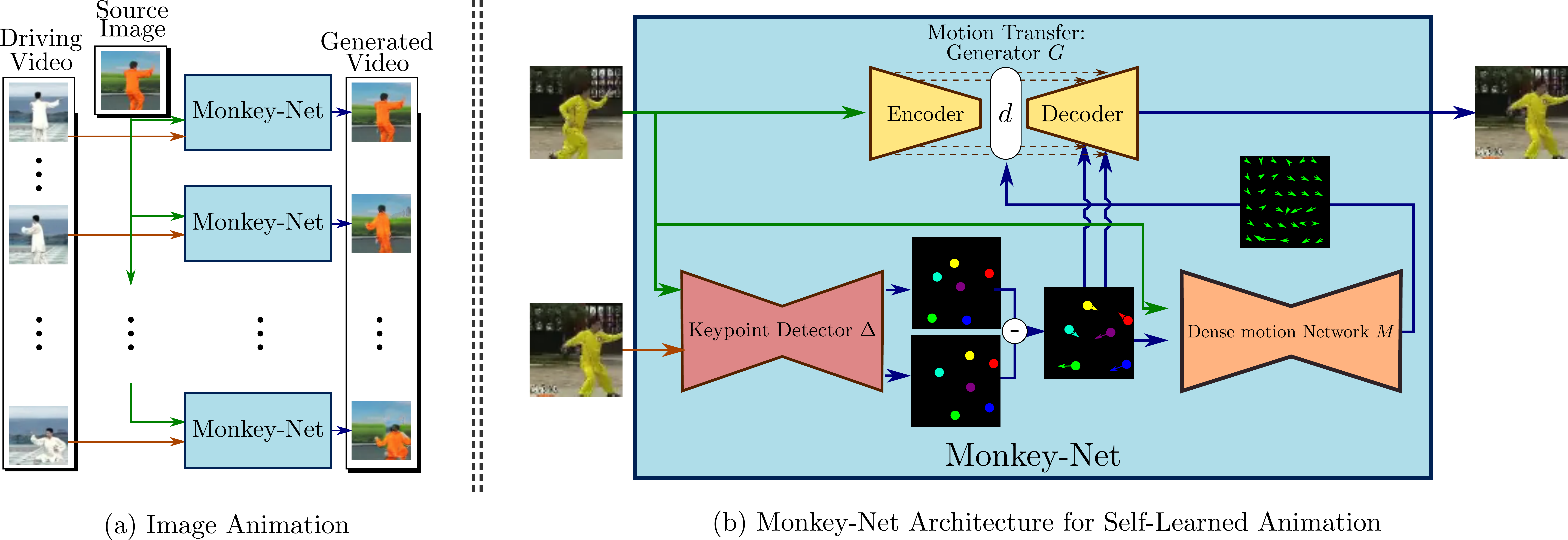}
\caption{A schematic representation of the proposed motion transfer framework for image animation. At testing time (Fig. (a)), the model generates a video with the object appearance of the source image but with the motion from the driving video. Monkey-Net (Fig. (b)) is composed of three networks: a motion-specific keypoint detector $\netDet$, a motion prediction network $M$ and an image generator $\netGen$. $\netGen$ reconstructs the image $\xvect'$ from the keypoint positions $\netDet(\xvect)$ and $\netDet(\xvect')$. The optical flow computed by $M$ is used by $\netGen$ to handle misalignments between $\xvect$ and $\xvect'$. The model is learned with a self-supervised learning scheme.}

\vspace{-0.5cm}
\label{fig:pipeline}

\end{figure*} 

\section{Monkey-Net}
\vspace{-0.2cm}
The architecture of the Monkey-Net is given in Fig.~\ref{fig:pipeline}. We now describe it in detail.

\label{sec:model}
\vspace{-0.2cm}
\subsection{Overview and Motivation}
\vspace{-0.2cm}
\label{sec:model-overview}
The objective of this work is to animate an object based on the motion of a similar object in a driving video. 
Our framework is articulated into three main modules (Fig.~\ref{fig:pipeline}). The first network, named Keypoint Detector, takes as input the source image and a frame from the driving video and automatically extracts sparse keypoints. The output of this module is then fed to a Dense Motion prediction network, which translates the sparse keypoints into 
motion heatmaps. The third module, the Motion Transfer network, receives as input the source image and the dense motion heatmap and recombines them 
producing a target frame. 

The output video is generated frame-by-frame as illustrated in Fig.~\ref{fig:pipeline}.a. At time $t$, 
the Monkey-Net uses the source image and the $t^{th}$ frame from the driving video. 
In order to train a Monkey-Net one just needs a dataset consisting of videos of objects of interest. No specific labels, such as keypoint annotations, are required. The learning process is fully self-supervised. Therefore, at test time, in order to generate a video sequence, the generator requires only a static input image and a motion descriptor from the driving sequence.
Inspired by recent studies on unsupervised landmark discovery for learning image representations~\cite{jakabunsupervised,Zhang_2018_CVPR}, we formulate the problem of learning a motion representation as an unsupervised motion-specific keypoint detection task. Indeed, the keypoints locations differences between two frames can be seen as a compact motion representation. In this way, our model generates a video by modifying the input image according to the landmarks extracted from the driving frames. Using a Monkey-Net at inference time is detailed in Sec.~\ref{sec:test}.


The Monkey-Net architecture is illustrated in Fig.~\ref{fig:pipeline}.b.
Let $\xvect$ and $\xvect'\in {\cal X}$ be two frames of size $H\times W$ extracted from the same video. The $H\times W$ lattice is denoted by $\mathcal{U}$. 
Inspired by \cite{jakabunsupervised}, we jointly learn a keypoint detector $\netDet$ together with a generator network $\netGen$ according to the following objective: $\netGen$ should be able to reconstruct $\xvect'$ from the keypoint locations $\netDet(\xvect)\in \mathcal{U}$, $\netDet(\xvect') \in \mathcal{U}$, and $\xvect$.
In this formulation, the motion between $\xvect$ and $\xvect'$ is implicitly modeled. 
To deal with large motions, we aim to learn keypoints that describe motion as well as the object geometry. To this end, we add a third network $M$ that estimates the optical flow $\mathcal{F}\in\mathbb{R}^{H\times W\times 2}$ between $\xvect'$ and $\xvect$ from $\netDet(\xvect)$, $\netDet(\xvect')$ and $\xvect$.
The motivation for this is twofold. First, this forces the keypoint detector $\netDet$ to predict keypoint locations that capture not only the object structure but also its motion. To do so, the learned keypoints must be located especially on the object parts with high probability of motion. For instance, considering the human body, it is important to obtain keypoints on the extremities (as in feet or hands) in order to describe the body movements correctly, since these body-parts tend to move the most. Second, following common practises in conditional image generation, the generator $\netGen$ is implemented as an encoder-decoder composed of convolutional blocks \cite{pix2pix2016}. However, standard convolutional encoder-decoders are not designed to handle large pixel-to-pixel misalignment between the input and output images \cite{siarohin2018deformable, balakrishnansynthesizing, ganin2016deepwarp}. 
To this aim, we introduce a deformation module within the generator $\netGen$ that employs the estimated optical flow $\mathcal{F}$ in order to align the encoder features~with~$\xvect'$.

\vspace{-0.2cm}
\subsection{Unsupervised Keypoint Detection}
\vspace{-0.2cm}
\label{sec:keypoint}
In this section, we detail the structure employed for unsupervised keypoint detection.
First, we employ a standard U-Net architecture that, from the input image, estimates $K$ heatmaps $H_k\in[0,1]^{H\times W}$, one for each keypoint. We employ softmax activations for the last layer of the decoder in order to obtain heatmaps that can be interpreted as detection confidence map for each keypoint.
An encoder-decoder architecture is used here since it has shown good performance for keypoints localization \cite{Bulat_2017_ICCV, robinson2019laplace}. 

To model the keypoint location confidence, we fit a Gaussian on each detection confidence map. Modeling the landmark location by a Gaussian instead of using directly the complete heatmap $H_k$ acts as a bottle-neck layer, and therefore allows the model to learn landmarks in an indirect way. The expected keypoint coordinates $\hvect_k\in\mathbb{R}$ and its covariance $\Sigma_k$ are estimated according to:
\begin{equation}
\hvect_k=\sum_{p\in\mathcal{U}}H_k[p]p;\; \Sigma_k=\sum_{p\in\mathcal{U}} H_k[p](p-\hvect_k)(p-\hvect_k)^\top\label{eq:cov}\\
\end{equation}
The intuition behind the use of keypoint covariances is that they can capture not only the location of a keypoint but also its orientation. Again considering the example of the human body: in the case of the legs, the covariance may capture their orientation. 
Finally, we encode the keypoint distributions as heatmaps $H^i_k\in[0,1]^{H\times W}$, such that they can be used as inputs to the generator and to the motion networks. Indeed, the advantage of using a heatmap representation, rather than considering directly the 2D coordinates $\hvect_k$, is that heatmaps are compatible with the use of convolutional neural networks.
Formally, we employ the following Gaussian-like function:
\begin{equation}
\forall p \in \mathcal{U},H_k(\mathbf{p}) = \frac{1}{\alpha} exp\left(-(\mathbf{p}- \hvect_k)\Sigma_k^{-1}(\mathbf{p}-\hvect_k)\right)
\end{equation}
where $\alpha$ is normalization constant.
This process is applied independently on $\xvect$ and $\xvect'$ leading to two sets of $K$ keypoints heatmaps $H=\{H_k\}_{k=1..K}$ and $H'=\{H_k'\}_{k=1..K}$.

\vspace{-0.2cm}
\subsection{Generator Network with Deformation Module}
\vspace{-0.2cm}
\label{sec:gen}

In this section, we detail how we reconstruct the target frame $\xvect'$ from $\xvect$, $\netDet(\xvect)=H$ and $\netDet(\xvect')=H'$. First we employ a standard convolutional encoder composed of a sequence of convolutions and average pooling layers in order to encode the object appearance in $\xvect$. Let $\xivect_r\in\mathbb{R}^{H_r\times W_r\times C_r} $ denote the output of the $r^{th}$ block of the encoder network ($1\leq r\leq R$). The architecture of this generator network is also based on the U-Net architecture \cite{ronneberger2015u} in order to obtain better details in the generated image. Motivated by \cite{siarohin2018deformable}, where it was shown that a standard U-net cannot handle large pixel-to-pixel misalignment between the input and the output images, we propose using a deformation module to align the features of the encoder with the output images. Contrary to \cite{siarohin2018deformable} that defines an affine transformation for each human body part in order to compute the feature deformation, we propose a deformation module that can be used on any object. In particular, we propose employing the optical flow $\mathcal{F}$ to align the features $\xivect_r$ with  $\xvect'$. 
The deformation employs a warping function $f_w(\cdot,\cdot)$ that warps the feature maps according to $\mathcal{F}$:
\begin{equation}
\xivect_r'=f_w(\xivect_r,\mathcal{F})\label{eq:wrap}
\end{equation}
This warping operation is implemented using a bilinear sampler, resulting in a fully differentiable model. Note that $\mathcal{F}$ is down-sampled to $H_r\times W_r$ via nearest neighbour interpolation when computing Eq. \eqref{eq:wrap}.
Nevertheless, because of the small receptive field of the bilinear sampling layer, encoding the motion only via the deformation module leads to optimization problems.
In order to facilitate network training,
we propose inputing the decoder the difference of the keypoint locations encoded as heatmaps $\dot{H}=H'-H$. Indeed, by providing $\dot{H}$ to the decoder, the reconstruction loss applied on the $\netGen$ outputs (see Sec.~\ref{sec:training}) is directly propagated to the keypoint detector $\netDet$ without going through $M$. In addition, the advantage of the heatmap difference representation is that it encodes both the locations and the motions of the keypoints. Similarly to $\mathcal{F}$, we compute $R$ tensors $\dot{H}_r$ by down-sampling $\dot{H}$ to $H_r\times W_r$. The two tensors $\dot{H}_r$ and $\xivect_r'$ are concatenated along the channel axis and are then treated as skip-connection tensors by the decoder.

\vspace{-0.2cm}
\subsection{From Sparse Keypoints to Dense Optical Flow}
\vspace{-0.2cm}
\label{sec:OF}

\begin{figure}[t]\centering
\includegraphics[width=0.99\columnwidth]{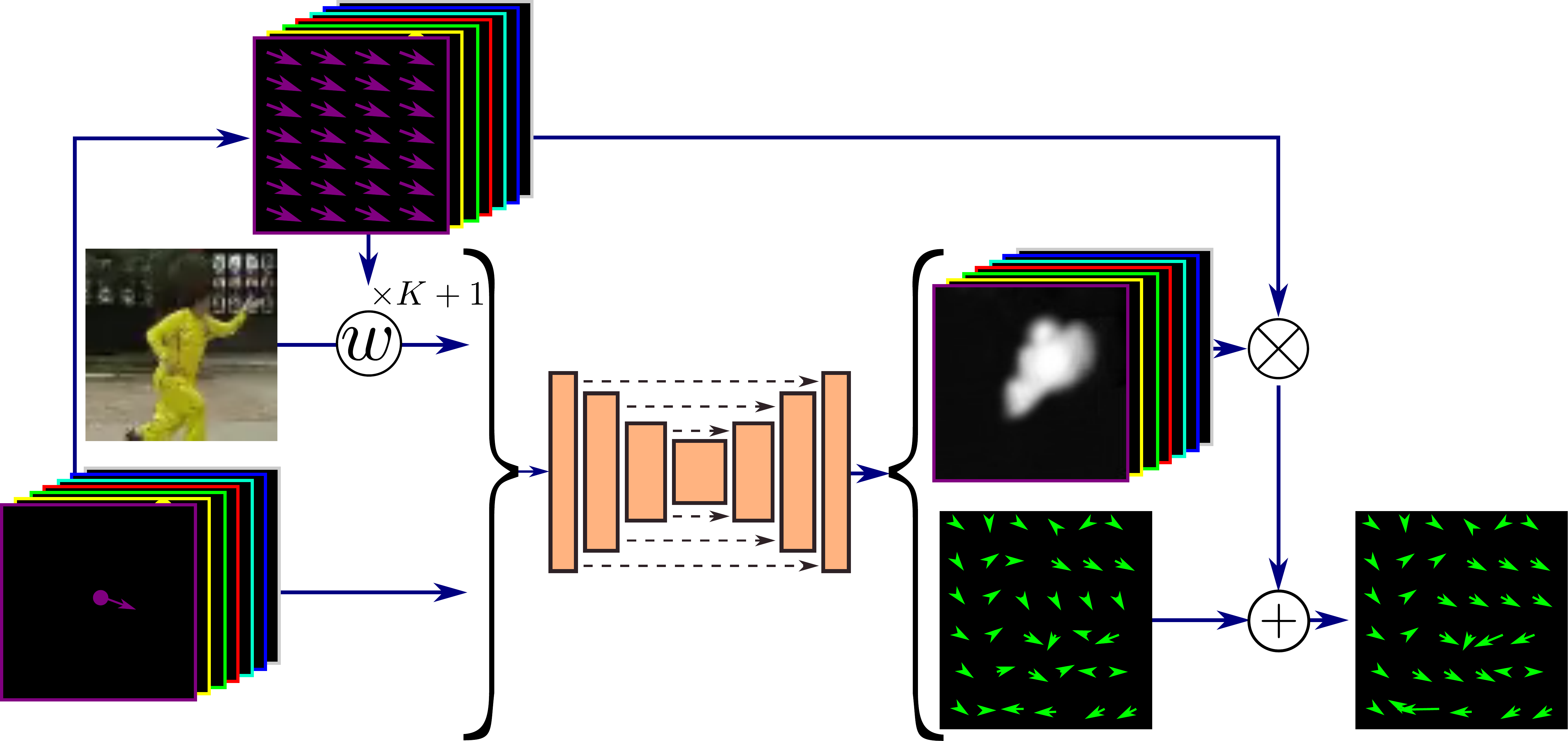}
\caption{A schematic representation of the adopted part-based model for optical flow estimation from sparse representation. From the appearance of the first frame and the keypoints motion, the network $M$ predicts a mask for each keypoint and the residual motion (see text for details).}

\label{fig:att-Unet}
\vspace{-0.5cm}
\end{figure}

In this section, we detail how we estimate the optical flow $\mathcal{F}$. The task of predicting a dense optical flow only from the displacement of a few keypoints and the appearance of the first frame is challenging. In order to facilitate the task of the network, we adopt a part base formulation. We make the assumption that each keypoint is located on an object part that is locally rigid. Thus, the task of computing the optical flow becomes simpler since, now, the problem consists in estimating masks $M_k\in \mathcal{R}^{H\times W}$ that segment the object in rigid parts corresponding to each keypoint. A first coarse estimation of the optical flow can be given by:
\begin{equation}
\mathcal{F}_\mathrm{coarse}=\sum_{k=1}^{K+1} M_k\otimes \rho(h_k)
\end{equation}
where $\otimes$ denotes the element-wise product and $\rho(\cdot)\in\mathcal{R}^{H\times W\times 2}$ is the operator that returns a tensor by repeating the input vector $H\times W$ times. Additionally, we employ one specific mask $M_{K + 1}$ without deformation (which corresponds to $\rho([0, 0])$) to capture the static background.
In addition to the masks $M_k$, the motion network $M$ also predicts the residual motion $\mathcal{F}_\mathrm{residual}$. The purpose of this residual motion field is to refine the coarse estimation by predicting non-rigid motion that cannot be modeled by the part-based approach. The final estimated optical flow is: $\mathcal{F}=\mathcal{F}_\mathrm{coarse}+\mathcal{F}_\mathrm{residual}$.

Concerning the inputs of the motion network, M takes two tensors, $\dot{H}$ and $\xvect$ corresponding respectively to the sparse motion and the appearance. However, we can observe that, similarly to the generator network, $M$ may suffer from the misalignment between the input $\xvect$ and the output $\mathcal{F}$. Indeed, $\mathcal{F}$ is aligned with $\xvect'$. To handle this problem, we use the warping operator $f_w$ according to the motion field of each keypoint $\rho(h_k)$, \eg $\xvect_k=f_w(\xvect,\rho(h_k))$. This solution provides images $\xvect_k$ that are locally aligned with $\mathcal{F}$ in the neighborhood of $h_k'$.
Finally, we concatenate $H'-H$, $\{\xvect_k\}_{k=1..K}$ and $\xvect$ along the channel axis and feed them into a standard U-Net network.
Similarly to the keypoint and the generator network, the use of U-Net architecture is motivated by the need of fine-grained details. 

\vspace{-0.2cm}
\subsection{Network Training}
\vspace{-0.2cm}
\label{sec:training}

We propose training the whole network in an end-to-end fashion.
As formulated in Sec.~\ref{sec:model-overview}, our loss ensures that $\xvect'$ is correctly reconstructed from $\netDet(\xvect)\in \mathcal{U}$, $\netDet(\xvect') \in \mathcal{U}$ and $\xvect$.
Following the recent advances in image generation, we combine an adversarial and the feature matching loss proposed in \cite{wang2017high} in order to learn to reconstruct $\xvect'$.
More precisely, we use a discriminator network $D$ that takes as input $H'$ concatenated with either the real image $\xvect'$ or the generated image $\xgen$. We employ the least-square GAN formulation \cite{mao2017least} leading to the two following losses used to train the discriminator and the generator:
\begin{equation}
\begin{split}
  \mathcal{L}^D_\mathrm{gan}(D) = &
  \mathbb{E}_{\xvect' \in {\cal X}}[(D(\xvect'\oplus H')-1)^2]  \\
  &+  \mathbb{E}_{(\xvect,\xvect') \in {\cal X}^2} [D(\xgen\oplus H'))^2]
 \end{split}
\notag
\end{equation}
\begin{equation}
\begin{split}
\mathcal{L}_\mathrm{gan}^{G}(G) = &\mathbb{E}_{(\xvect,\xvect')\in {\cal X}^2}[(D(\xgen\oplus H')-1)^2]
\end{split}
\label{eq:GAN-loss}
\end{equation}
where $\oplus$ denotes the concatenation along the channel axis. Note that in Eq \eqref{eq:GAN-loss}, the dependence on the trained parameters of $\netGen$, $M$, and $\netDet$ appears implicitly via $\xgen$. Note that we provide the keypoint locations $H'$ to the discriminator to help it to focus on moving parts and not on the background. However, when updating the generator, we do not propagate the discriminator loss gradient through $H'$ to avoid that the generator tends to fool the discriminator by generating meaningless keypoints.

The GAN loss is combined with a feature matching loss that encourages the output image $\xgen$  and $\xvect'$ to have similar feature representations. The feature representations employed to compute this loss are the intermediate layers of the discriminator $D$. The feature matching loss is given by:
\begin{equation}
\mathcal{L}_\mathrm{rec}=\mathbb{E}_{(\xvect,\xvect') }\left[\Vert D_i(\xgen \oplus H')-D_i(\xvect' \oplus H'))\Vert_1 \right] \label{eq:recLoss}
\end{equation}
where $D_i$ denotes the $i^{th}$-layer feature extractor of the discriminator $D$. $D_0$ denotes the discriminator input. The main advantage of the feature matching loss is that, differently from other perceptual losses, \cite{siarohin2018deformable,johnson2016perceptual}, it does not require the use of an external pre-trained network. 
Finally the overall loss is obtained by combining Eqs.~\eqref{eq:recLoss} and \eqref{eq:GAN-loss},  $\mathcal{L}_\mathrm{tot}=\lambda_\mathrm{rec}\mathcal{L}_\mathrm{rec}+ {\cal L}^G_\mathrm{gan}$.
In all our experiments, we chose  $\lambda_{rec}=10$ following \cite{wang2017high}. Additional details of our implementation are shown in the Supplementary Material A.

\vspace{-0.2cm}
\subsection{Generation Procedure}
\vspace{-0.2cm}
\label{sec:test}

At test time, our network receives a driving video and a source image. In order to generate the $t^{th}$ frame, $\Delta$ estimates the keypoint locations $h^s_k$ in the source image. Similarly, we estimate the keypoint locations $h^1_k$ and $h^t_k$ from  first and the $t^{th}$ frames of the driving video. Rather than generating a video from the absolute positions of the keypoints, the source image keypoints are transferred according to the relative difference between keypoints in the video. The keypoints in the generated frame are given by: 
\begin{equation}
{h_k^{s}}'=h^s_k+(h^t_k-h^1_k)
\label{eq:mot}
\end{equation}
The keypoints ${h_k^{s}}'$ and $h^s_k$ are then encoded as heatmaps using the covariance matrices estimated from the driving video, as described in Sec.~\ref{sec:keypoint}. Finally, the heatmaps are given to the dense motion and the generator networks together with the source image (see Secs.~\ref{sec:gen} and \ref{sec:OF}). Importantly, one limitation of transferring relative motion is that it cannot be applied to arbitrary source images. Indeed, if the driving video object is not roughly aligned with the source image object, Eq.~\eqref{eq:mot} may lead to absolute keypoint positions that are physically impossible for the considered object as illustrated in Supplementary Material C.1.
\vspace{-0.2cm}
\section{Experiments}
\label{Experiments}
\vspace{-0.2cm}

In this section, we present a in-depth evaluation on three problems, tested on three very different datasets and employing a large variety of metrics. 


\noindent
\textbf{Datasets.} The UvA-\emph{Nemo} dataset \cite{dibekliouglu2012you} is a facial dynamics analysis dataset composed of 1240 videos
We follow the same pre-processing as in \cite{wang2018every}. Specifically, faces are aligned using the OpenFace library \cite{amos2016openface} before re-sizing each frame to $64\times64$ pixels. Each video starts from a neutral expression and lasts 32 frames. As in \cite{wang2018every}, we use 1110 videos for training and 124 for evaluation. 

The \emph{Tai-Chi} dataset \cite{tulyakov2017mocogan} is composed of 4500 tai-chi video clips downloaded from YouTube. We use the data as pre-processed in \cite{tulyakov2017mocogan}. In particular, the frames are resized to $64\times64$ pixels. The videos are split into 3288 and 822 videos for training and testing respectively. The video length varies from 32 to 100 frames.

The \emph{BAIR} robot pushing dataset \cite{ebert2017self} contains videos
collected by a Sawyer robotic arm pushing a variety of objects over a table. It contains 40960 training and 256 test videos. Each video is $64\times64$ pixels and has 
30 frames. 

\noindent
\textbf{Evaluation Protocol.}\label{sec:pb}
Evaluating the results of image animation methods is a difficult task, since ground truth animations are not available. In addition, to the best of our knowledge, X2Face \cite{wiles2018x2face} is the only previous approach for data-driven model-free image animation.
For these two reasons, we evaluate our method also on two closely related tasks. As proposed in \cite{wiles2018x2face}, we first evaluate Monkey-Net on the task of video reconstruction. This consists in reconstructing the input video from a representation in which motion and content are decoupled. 
This task is a ``proxy'' task to image animation and it is only introduced for the purpose of quantitative comparison.
In our case, we combine the extracted keypoints $\Delta(x)$ of each frame and the first frame of the video to re-generate the input video. 
Second, we evaluate our approach on the problem of Image-to-Video translation. Introduced in \cite{vondrick2016generating}, this problem consists of generating a video from its first frame. Since our model is not directly designed for this task, we train a small recurrent neural network that predicts, from the keypoint coordinates in the first frame, the sequence of keypoint coordinates for the other 32 frames.
Additional details can be found in the Supplementary Material A.
Finally, we evaluate our model on 
image animation. In all experiments we use K=10.

\begin{table}[t]
\small
    \resizebox{\columnwidth}{!}{
    \begin{tabular}{c|ccc|ccc|c}
    \toprule
        & \multicolumn{3}{c|}{\emph{Tai-Chi}} & \multicolumn{3}{c|}{Nemo} & \multicolumn{1}{c}{Bair} \\
        & $\mathcal{L}_1$  & ({AKD}, {MKR}) & {AED} & $\mathcal{L}_1$ & {AKD} & {AED} & $\mathcal{L}_1$  \\
        \midrule
        X2Face & 0.068 &  (4.50, 35.7\%)  & 0.27 & 0.022 & 0.47 & 0.140 & 0.069 \\
        Ours   & \bf0.050 & (\bf2.53, \bf17.4\%) & \bf0.21 & \bf0.017 & \bf0.37 & \bf0.072 & \bf0.025 \\
        \bottomrule
    \end{tabular}
    }
\captionof{table}{Video reconstruction comparisons}
    \label{tab:recSota}
\vspace{-0.5cm}
\end{table}

\noindent
\textbf{Metrics.}
 In our experiments, we adopt several metrics in order to provide an in-depth comparison with other methods. 
 We employ the following metrics. 
 \begin{itemize}[noitemsep,topsep=0pt,wide=0pt]
 \item $\mathcal{L}_1$. In the case of the video reconstruction task where the ground truth video is available, we compare the average $\mathcal{L}_1$ distance between pixel values of the ground truth and the generated video frames. 
 \item AKD. For the  \emph{Tai-Chi} and \emph{Nemo} datasets, we employ external keypoint detectors in order to evaluate whether the motion of the generated video matches the ground truth video motion. For the \emph{Tai-Chi} dataset, we employ the human-pose estimator in \cite{cao2017realtime}. For the \emph{Nemo} dataset we use the facial landmark detector of \cite{Bulat_2017_ICCV}.
We compute these keypoints for each frame of the ground truth and the generated videos. From these externally computed keypoints, we deduce the \emph{Average Keypoint Distance} (AKD), \ie the average distance between the detected keypoints of the ground truth and the generated video. 
\item MKR. In the case of the \emph{Tai-Chi} dataset, the human-pose estimator returns also a binary label for each keypoint indicating whether the keypoints were successfully detected. Therefore, we also report the \emph{Missing Keypoint Rate} (MKR) that is the percentage of keypoints that are detected in the ground truth frame but not in the generated one. This metric evaluates the appearance quality of each video frame. 
\item AED. We compute the feature-based metric employed in \cite{esser2018variational} that consists in computing the \emph{Average Euclidean Distance (AED)} between a feature representation of the ground truth and the generated video frames. The feature embedding is chosen such that the metric evaluates how well the identity is preserved. More precisely, we use a network trained for facial identification \cite{amos2016openface} for \emph{Nemo} and a network trained for person re-id \cite{hermans2017defense} for \emph{Tai-Chi}.  
\item FID. When dealing with Image-to-video translation, we complete our evaluation with the \textit{Frechet Inception Distance} \cite{heusel2017gans} (FID)  in order to evaluate the quality of individual frames. 
 \end{itemize}
Furthermore, we conduct a user study for both the Image-to-Video translation 
and the image animation tasks (see Sec.~\ref{sec:us}). 


\vspace{-0.2cm}
\subsection{Ablation Study}
\vspace{-0.2cm}
In this section, we present an ablation study to empirically measure the impact of each part of our proposal on the  performance. First, we describe the methods obtained by ``amputating'' key parts of the model described in Sec.~\ref{sec:model-overview}: (i) \emph{No $\mathcal{F}$} - the dense optical flow network $M$ is not used; 
(ii) \emph{No $\mathcal{F}_\mathrm{coarse}$} - in the optical flow network $M$, we do not use the part based-approach; 
(iii) \emph{No $\mathcal{F}_{residual}$} - in the Optical Flow network $M$, we do not use $\mathcal{F}_\mathrm{residual}$; 
(iv) \emph{No $\Sigma_k$} - we do not estimate the covariance matrices $\Sigma_K$ in the keypoint detector $\Delta$  and the variance is set to $\Sigma_k=0.01$ as in \cite{jakabunsupervised};
(v) the source image is not given to the motion network $M$, $M$ estimates the dense optical flow only from the keypoint location differences; 
(vi) \emph{Full} denotes the full model as described in Sec.~\ref{sec:model}.

\begin{table}[t]
    \centering
    \small
    \begin{tabular}{cccc}
        \toprule
        \multicolumn{4}{c}{\emph{Tai-Chi}} \\
        \midrule
        & $\mathcal{L}_1$ & ({AKD}, {MKR}) & {AED} \\
        \midrule
        \emph{No $\mathcal{F}$} & 0.057 & (3.11, 23.8\%) & 0.24 \\
        \emph{No $\mathcal{F}_\mathrm{residual}$} & 0.051 & (2.81, 18.0\%) & 0.22 \\ 
        \emph{No $\mathcal{F}_{coarse}$}  & 0.052 & (2.75, 19.7\%) & 0.22 \\
        \emph{No $\Sigma_k$} & 0.054 & (2.86, 20.6\%)  & 0.23 \\
        \emph{No $\xvect$} & 0.051 & (2.71, 19.3\%) & \bf0.21 \\
        
        \emph{Full} & \bf0.050 & (\bf2.53, \bf17.4\%)& \bf0.21 \\
        \bottomrule
    \end{tabular}
    \captionof{table}{Video reconstruction ablation study \emph{TaiChi}.}
    \label{tab:ablations}
    \vspace{-0.5cm}
\end{table}
\begin{figure}[t]
  \centering
  \setlength\tabcolsep{0.5pt}
\resizebox{0.85\columnwidth}{!}{\begin{tabular}{>{\centering\arraybackslash}m{1.3cm}ccccc}
\\


\vspace{-1cm}Real $\xvect'$  & \includegraphics[width=0.14\columnwidth]{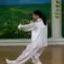}&
\includegraphics[width=0.14\columnwidth]{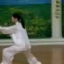}&
\includegraphics[width=0.14\columnwidth]{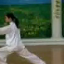}&
\includegraphics[width=0.14\columnwidth]{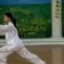}&
\includegraphics[width=0.14\columnwidth]{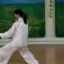}
\\
\vspace{-1cm}\emph{No $\mathcal{F}$} & \includegraphics[width=0.14\columnwidth]{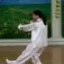}&
\includegraphics[width=0.14\columnwidth]{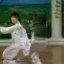}&
\includegraphics[width=0.14\columnwidth]{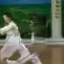}&
\includegraphics[width=0.14\columnwidth]{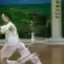}&
\includegraphics[width=0.14\columnwidth]{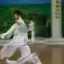}
\\
\vspace{-1cm}\emph{No}\newline $\mathcal{F}_\mathrm{coarse}$ & \includegraphics[width=0.14\columnwidth]{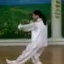}&
\includegraphics[width=0.14\columnwidth]{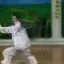}&
\includegraphics[width=0.14\columnwidth]{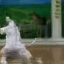}&
\includegraphics[width=0.14\columnwidth]{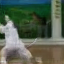}&
\includegraphics[width=0.14\columnwidth]{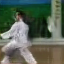}
\\
\vspace{-1cm}\emph{Full} & \includegraphics[width=0.14\columnwidth]{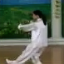}&
\includegraphics[width=0.14\columnwidth]{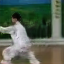}&
\includegraphics[width=0.14\columnwidth]{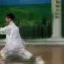}&
\includegraphics[width=0.14\columnwidth]{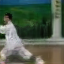}&
\includegraphics[width=0.14\columnwidth]{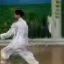}
\end{tabular}}
  \caption{Qualitative ablation evaluation of video reconstruction on \emph{Tai-Chi}.}
\label{fig:ablations}
\vspace{-0.5cm}
\end{figure}

In Tab.~\ref{tab:ablations}, we report the quantitative evaluation. 
We first observe that our full model outperforms the baseline method without deformation. This trend is observed according to all the metrics. This illustrates the benefit of deforming the features maps according to the estimated motion. 
Moreover, we note that \emph{No} $\mathcal{F}_{coarse}$ and \emph{No} $\mathcal{F}_\mathrm{residual}$ both perform worse than when using the full optical flow network. This illustrates that $\mathcal{F}_{coarse}$ and $\mathcal{F}_\mathrm{residual}$ alone are not able to estimate dense motion accurately.
 A possible explanation is that $\mathcal{F}_{coarse}$ cannot estimate non rigid motions and that $\mathcal{F}_\mathrm{residual}$, on the other hand, fails in predicting the optical flow in the presence of large motion. The qualitative results shown in Fig.~\ref{fig:ablations} confirm this analysis.
Furthermore, we observe a drop in performance when covariance matrices are replaced with static diagonal matrices. 
This shows the benefit of encoding more information 
when dealing with videos with complex and large motion, as in the case of the \emph{TaiChi} dataset.
Finally, we observe that if the appearance is not provided to the deformation network $M$, the video reconstruction performance is slightly lower.  




\begin{table}[t]
    \centering
     \resizebox{0.9\textwidth}{!}{
     \begin{minipage}{\textwidth}
     \hspace{1.2cm}\begin{tabular}{cccc}
    
     \toprule
        \multicolumn{4}{c}{\emph{Tai-Chi}} \\
        \midrule
         & FID & AED & MKR \\
    \midrule    
        MoCoGAN \cite{tulyakov2017mocogan} &  54.83 & 0.27 & 46.2\%  \\
        Ours  & \bf19.75 & \bf0.17 & \bf 30.3\%   \\
                 \bottomrule
                 \end{tabular}
                 \vspace{0.2cm}
                 
\begin{minipage}[b]{0.3\textwidth} 
\begin{tabular}{ccc}
\toprule    
         \multicolumn{3}{c}{\emph{Nemo}} \\
         \midrule
      & FID & AED   \\
        \midrule
         MoCoGAN \cite{tulyakov2017mocogan} & 51.50 & 0.33 \\
        CMM-Net \cite{wang2018every} & 27.27 & 0.13 \\
        Ours   & \bf11.97 & \bf0.12  \\
        \bottomrule
         \end{tabular}
         \end{minipage}
          \begin{minipage}[b]{0.3\textwidth}
\begin{tabular}{cc}
        \toprule
       \multicolumn{2}{c}{\emph{Bair}} \\
       \midrule
       & FID    \\
       \midrule
        MoCoGAN \cite{tulyakov2017mocogan} & 244.00\\
        SV2P \cite{babaeizadeh2017stochastic}   & 57.90 \\
        Ours    &  \bf23.20\\
        \bottomrule
    \end{tabular}  
    \end{minipage}
        \end{minipage}
}   
     \caption{Image-to-video translation comparisons.}
    \label{tab:img2vid}
\vspace{-0.5cm}
\end{table}



\begin{figure*}[t]
  \centering
  \setlength\tabcolsep{0.5pt}
\resizebox{0.33\linewidth}{!}{\begin{tabular}{cccccc}
  \includegraphics[height=0.14\columnwidth]{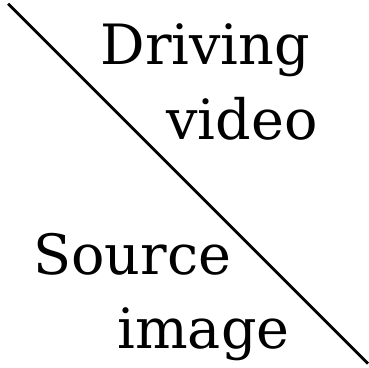}&
 \includegraphics[height=0.14\columnwidth]{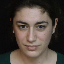}&
  \includegraphics[height=0.14\columnwidth]{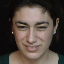}&
\includegraphics[height=0.14\columnwidth]{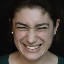}&
\includegraphics[height=0.14\columnwidth]{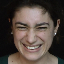}
\\
\includegraphics[height=0.14\columnwidth]{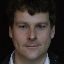}&
\includegraphics[height=0.14\columnwidth]{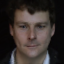}&
\includegraphics[height=0.14\columnwidth]{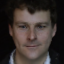}&
\includegraphics[height=0.14\columnwidth]{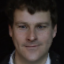}&
\includegraphics[height=0.14\columnwidth]{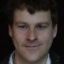}
\\
\includegraphics[height=0.14\columnwidth]{figures/nemo-transfer/001app001.png}&
\includegraphics[height=0.14\columnwidth]{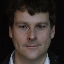}&
\includegraphics[height=0.14\columnwidth]{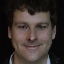}&
\includegraphics[height=0.14\columnwidth]{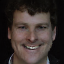}&
\includegraphics[height=0.14\columnwidth]{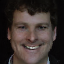}
\end{tabular}}
\resizebox{0.32\linewidth}{!}{\begin{tabular}{cccccc}
\includegraphics[height=0.14\columnwidth]{figures/tablecap.pdf}&
 \includegraphics[height=0.14\columnwidth]{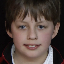}&
  \includegraphics[height=0.14\columnwidth]{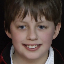}&
\includegraphics[height=0.14\columnwidth]{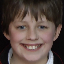}&
\includegraphics[height=0.14\columnwidth]{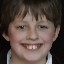}
\\
\includegraphics[height=0.14\columnwidth]{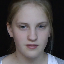}&
\includegraphics[height=0.14\columnwidth]{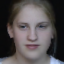}&
\includegraphics[height=0.14\columnwidth]{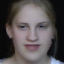}&
\includegraphics[height=0.14\columnwidth]{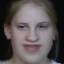}&
\includegraphics[height=0.14\columnwidth]{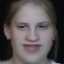}
\\
\includegraphics[height=0.14\columnwidth]{figures/nemo-transfer/001app002.png}&
\includegraphics[height=0.14\columnwidth]{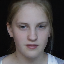}&
\includegraphics[height=0.14\columnwidth]{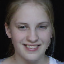}&
\includegraphics[height=0.14\columnwidth]{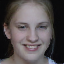}&
\includegraphics[height=0.14\columnwidth]{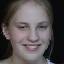}
\end{tabular}}
\resizebox{0.32\linewidth}{!}{\begin{tabular}{cccccc}
\includegraphics[height=0.14\columnwidth]{figures/tablecap.pdf}&
\includegraphics[height=0.14\columnwidth]{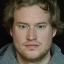}&
  \includegraphics[height=0.14\columnwidth]{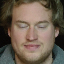}&
\includegraphics[height=0.14\columnwidth]{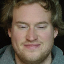}&
\includegraphics[height=0.14\columnwidth]{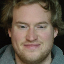}
\\
\includegraphics[height=0.14\columnwidth]{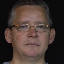}&
\includegraphics[height=0.14\columnwidth]{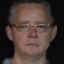}&
\includegraphics[height=0.14\columnwidth]{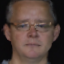}&
\includegraphics[height=0.14\columnwidth]{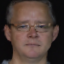}&
\includegraphics[height=0.14\columnwidth]{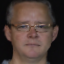}
\\
\includegraphics[height=0.14\columnwidth]{figures/nemo-transfer/001app003.png}&
\includegraphics[height=0.14\columnwidth]{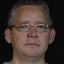}&
\includegraphics[height=0.14\columnwidth]{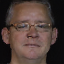}&
\includegraphics[height=0.14\columnwidth]{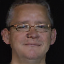}&
\includegraphics[height=0.14\columnwidth]{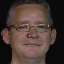}
\\
\end{tabular}
}
\vspace{-0.2cm}
  \caption{Qualitative results for image animation on the Nemo dataset: X2face (2-nd row) against our method (3-rd row).
 }
 \vspace{-0.2cm}
\label{fig:nemo-transfer-main}
\end{figure*}
\begin{figure*}[h!]
  \centering
  \setlength\tabcolsep{0.5pt}
\resizebox{0.33\linewidth}{!}{\begin{tabular}{cccccc}
\includegraphics[height=0.14\columnwidth]{figures/tablecap.pdf}&
\includegraphics[height=0.14\columnwidth]{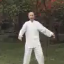}&
  \includegraphics[height=0.14\columnwidth]{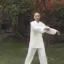}&
\includegraphics[height=0.14\columnwidth]{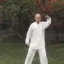}&
\includegraphics[height=0.14\columnwidth]{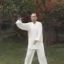}
\\
\includegraphics[height=0.14\columnwidth]{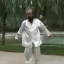}&
\includegraphics[height=0.14\columnwidth]{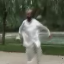}&
\includegraphics[height=0.14\columnwidth]{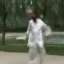}&
\includegraphics[height=0.14\columnwidth]{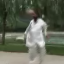}&
\includegraphics[height=0.14\columnwidth]{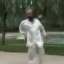}
\\
\includegraphics[height=0.14\columnwidth]{figures/taichi-transfer/001app001.png}&
\includegraphics[height=0.14\columnwidth]{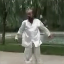}&
\includegraphics[height=0.14\columnwidth]{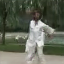}&
\includegraphics[height=0.14\columnwidth]{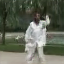}&
\includegraphics[height=0.14\columnwidth]{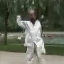}
\end{tabular}}
\resizebox{0.33\linewidth}{!}{\begin{tabular}{cccccc}
\includegraphics[height=0.14\columnwidth]{figures/tablecap.pdf}&
\includegraphics[height=0.14\columnwidth]{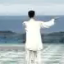}&
  \includegraphics[height=0.14\columnwidth]{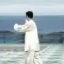}&
\includegraphics[height=0.14\columnwidth]{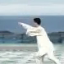}&
\includegraphics[height=0.14\columnwidth]{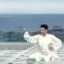}
\\
\includegraphics[height=0.14\columnwidth]{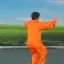}&
\includegraphics[height=0.14\columnwidth]{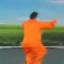}&
\includegraphics[height=0.14\columnwidth]{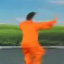}&
\includegraphics[height=0.14\columnwidth]{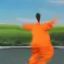}&
\includegraphics[height=0.14\columnwidth]{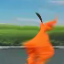}
\\
\includegraphics[height=0.14\columnwidth]{figures/taichi-transfer/001app002.png}&
\includegraphics[height=0.14\columnwidth]{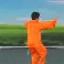}&
\includegraphics[height=0.14\columnwidth]{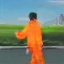}&
\includegraphics[height=0.14\columnwidth]{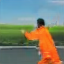}&
\includegraphics[height=0.14\columnwidth]{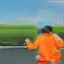}
\end{tabular}}
\resizebox{0.33\linewidth}{!}{\begin{tabular}{cccccc}
\includegraphics[height=0.14\columnwidth]{figures/tablecap.pdf}&
\includegraphics[height=0.14\columnwidth]{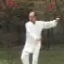}&
  \includegraphics[height=0.14\columnwidth]{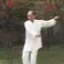}&
\includegraphics[height=0.14\columnwidth]{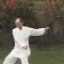}&
\includegraphics[height=0.14\columnwidth]{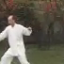}
\\
\includegraphics[height=0.14\columnwidth]{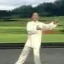}&
\includegraphics[height=0.14\columnwidth]{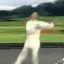}&
\includegraphics[height=0.14\columnwidth]{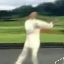}&
\includegraphics[height=0.14\columnwidth]{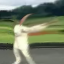}&
\includegraphics[height=0.14\columnwidth]{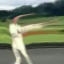}
\\
\includegraphics[height=0.14\columnwidth]{figures/taichi-transfer/001app003.png}&
\includegraphics[height=0.14\columnwidth]{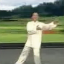}&
\includegraphics[height=0.14\columnwidth]{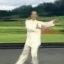}&
\includegraphics[height=0.14\columnwidth]{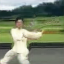}&
\includegraphics[height=0.14\columnwidth]{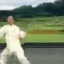}
\end{tabular}}
 \vspace{-0.3cm}
  \caption{Qualitative results for image animation on the \emph{Tai-Chi} dataset: X2face (2-nd row) against our method (3-rd row) 
  }
\label{fig:taichi-transfer-main}
\vspace{-0.5cm}
\end{figure*}

\vspace{-0.2cm}
\subsection{Comparison with Previous Works}
\label{sec:sota}
\vspace{-0.2cm}

\textbf{Video Reconstruction.}
First, we compare our results with the X2Face model \cite{wiles2018x2face} that is closely related to our proposal. Note that this comparison can be done since we employ image and motion representation of similar dimension. In our case, each video frame is reconstructed from the source image and 10 landmarks, each one represented by 5 numbers (two for the location and three for the symmetric covariance matrix), leading to a motion representation of dimension 50. For X2face, motion is encoded into a driving vector of dimension 128. 
The quantitative comparison is reported in Tab.~\ref{tab:recSota}. Our approach outperforms X2face, according the all the metrics and on all the evaluated datasets. This confirms
that encoding motion via motion-specific keypoints leads to a compact but rich representation.

\textbf{Image-to-Video Translation:}
In Tab.~\ref{tab:img2vid} we compare with the state of the art Image-to-Video translation methods: two unsupervised methods MoCoGAN~\cite{tulyakov2017mocogan} and SV2P~\cite{babaeizadeh2017stochastic}, and CMM-Net which is based on keypoints~\cite{wang2018every}. CMM-Net is evaluated only on \emph{Nemo} since it requires facial landmarks. We report results SV2P on the \emph{Bair} dataset as in \cite{babaeizadeh2017stochastic}.
We can observe that our method clearly outperforms the three methods for all the metrics. This quantitative evaluation is confirmed by the qualitative evaluation presented in the Supplementary material C.3. 
In the case of MoCoGAN,
we observe that the {AED} score is much higher than the two other methods. Since {AED} measures how well the identity is preserved, these results confirm that, despite the realism of the video generated by MoCoGAN, the identity and the person-specific details are not well preserved.
A possible explanation is that MoCoGAN is based on a feature embedding in a vector, which does not capture spatial information as well as the keypoints. The method in \cite{wang2018every} initially produces a realistic video and preserves the identity, but the lower performance can be explained by the apparition of visual artifacts in the presence of large motion (see the Supplementary material C.3 for visual examples). Conversely, our method both preserves the person identity and performs well even under large spatial deformations.

\vspace{-0.1cm}
\noindent\textbf{Image Animation.} 
In Fig.~\ref{fig:nemo-transfer-main}, we  compare our method with X2Face~\cite{wiles2018x2face} on the \emph{Nemo} dataset. We note that our method generates more realistic smiles on the three randomly selected samples despite the fact that the XFace model is specifically designed for faces. Moreover, the benefit of transferring the relative motion over absolute locations can be clearly observed in Fig.~\ref{fig:nemo-transfer-main} (column 2). When absolute locations are transferred, the source image inherits the face proportion from the driving video, resulting in a face with larger cheeks.
In Fig.~\ref{fig:taichi-transfer-main}, we compare our method with X2Face on the \emph{Tai-Chi} dataset. X2Face~\cite{wiles2018x2face} fails to consider each body-part independently and, consequently, warps the body in such a way that its center of mass matches the center of mass in the driving video. Conversely, our method successfully generates plausible motion sequences that match the driving videos. Concerning the \emph{Bair} dataset, exemplar videos are shown in the Supplementary material C.3.
The results are well in line with those obtained on the two other datasets. 

\begin{table}[t]
    \centering
    \small
    \begin{tabular}{c|c|c}
    \toprule
         \multicolumn{1}{c|}{\emph{Tai-Chi}} & \multicolumn{1}{c|}{\emph{Nemo}} & \multicolumn{1}{c}{\emph{Bair}} \\
        \midrule
         85.0\%  & 79.2\% & 90.8\% \\
        \bottomrule
    \end{tabular}
    \captionof{table}{User study results on image animation. 
    Proportion of times our approach is preferred over X2face~\cite{wiles2018x2face}.}
    \label{tab:transfer-user}
    \vspace{-0.5cm}
\end{table}


\vspace{-0.2cm}
\subsection{User Evaluation}
\vspace{-0.2cm}

\label{sec:us}

In order to further consolidate the quantitative and qualitative evaluations, we performed user studies for both the Image-to-Video translation (see the Supplementary Material C.3) and the image animation problems using Amazon Mechanical Turk.

For the image animation problem, our model is again compared with X2face~\cite{wiles2018x2face} according to the following protocol: we randomly select 50 pairs of videos where objects in the first frame have a similar pose. Three videos are shown to the user: one is the driving video (reference) and 2 videos from our method and X2Face. The users are given the following instructions: \textit{Select the video that better corresponds to the animation in the reference video}. 
We collected annotations for each video from 10 different users
The results are presented in Tab.~\ref{tab:transfer-user}. 
Our generated videos are preferred over X2Face videos in almost more than 80\% of the times for all the datasets. 
Again, we observe that the preference toward our approach is higher on the two datasets which correspond to large motion patterns.

\vspace{-0.2cm}
\section{Conclusion}
\vspace{-0.2cm}

We introduced a novel deep learning approach for image animation. Via the use of motion-specific keypoints, previously learned following a self-supervised approach, our model can animate images of arbitrary objects according to the motion given by a driving video. Our experiments, considering both automatically computed metrics and human judgments, demonstrate that the proposed method outperforms previous work on unsupervised image animation. Moreover, we show that with little adaptation our method can perform Image-to-Video translation. 
In future work, we plan to extend our framework to handle multiple objects and investigate other strategies for motion embedding.


\vspace{-0.4cm}
\subsubsection*{Acknowledgments}
\vspace{-0.4cm}
This work was carried out under the \enquote{Vision and Learning joint Laboratory} between FBK and UNITN.


{\small
\bibliographystyle{ieee}
\bibliography{egbib}
}
\clearpage

\renewcommand{\thesection}{\Alph{section}}
\setcounter{section}{0}

In this supplementary material, we provide implementation details (Sec.~\ref{sec:impl}), introduce a new dataset (Sec.~\ref{sec:mgif}) and report additional experimental results (Sec.~\ref{sec:addQual}). Additionally we provide a video file with further qualitative examples.

\section{Implementation details}
\label{sec:impl}
As described in Sec.~3, each module employs a U-Net architecture. We use the exact same architecture for all the networks. More specifically each block of each of the encoder consists of a $3 \times 3$ convolution, batch normalization \cite{ioffe2015batch}, ReLU and average pooling. The first convolution layers have 32 filters and each subsequent convolution doubles the number of filters. Each encoder is composed of a total of 5 blocks. The decoder blocks have similar structure: $3 \times 3$ convolution, batch normalization  and ReLU followed by nearest neighbour up-sampling.  The first block of the decoder has 512 filters. Each consequent block has the reduced number of filters by a factor of 2. 

As described in Sec.~3.2, the keypoint detector $\Delta$ produces $K$ heatmaps followed  by softmax. In particular, we employ softmax activations with 0.1 temperature. Indeed, thanks to the use of a low temperature for softmax, we obtain sharper heatmaps and avoid uniform heatmaps that would lead to keypoints constantly located in the image center.

For $G$, we employ 4 additional Residual Blocks \cite{he2016deep} in order to remove possible warping artifacts produces by $M$. The output of $G$ is a 3 channel feature map followed by the sigmoid. We use the discriminator architecture described in \cite{wang2017high}.

The framework is trained for $T$ epochs where $T$ equals 250, 500 and 10 for \emph{Tai-Chi}, \emph{Nemo} and \emph{Bair} respectively. Epoch involves training the network on 2 randomly sampled frames from each training video. We use the Adam optimizer~\cite{kingma2014adam} with learning rate 2e-4 and then with learning rate 2e-5 for another $\frac{T}{2}$  epochs. 

As explained in Sec.~4.2, for \emph{Image-to-Video} translation, we employ a single-layer GRU network in order to predict the keypoint sequence used to generate the video. This recurrent network \cite{chung2014empirical} has 1024 hidden units and is trained via $\mathcal{L}_1$ minimization.

\section{MGif dataset}
\label{sec:mgif}
We collected an additional dataset of videos containing movements of different cartoon animals. Each video is a moving \emph{gif} file. Therefore, we called this new dataset \emph{MGif}. The dataset consists of 1000 videos, we used 900 videos for training and 100 for evaluation. Each video has size $128 \times 128$ and contains from 5 to 100 frames. The dataset is particularly challenging because of the high appearance variation and motion diversity. Note that in the experiments on this dataset, we use absolute keypoint locations from the driving video instead of the relative keypoint motion detailed in Sec.~3.6.

\section{Additional experimental results}
\label{sec:addQual}

In this section, we report additional results.
In Sec.~\ref{sec:alignment} we visually motivate our alignment assumption, in Sec.~\ref{sec:abla} we complete the ablation study and, in Secs.~\ref{sec:i2v} and \ref{sec:ia}, we report qualitative results for both the image-to-video and image animation problems. Finally, in Sec.~\ref{sec:kv}, we visualize the keypoint predicted by our self-supervised approach.

\subsection{Explanation of alignment assumption}
\label{sec:alignment}
\begin{figure}[h!]
  \centering
\includegraphics[width=0.9\columnwidth]{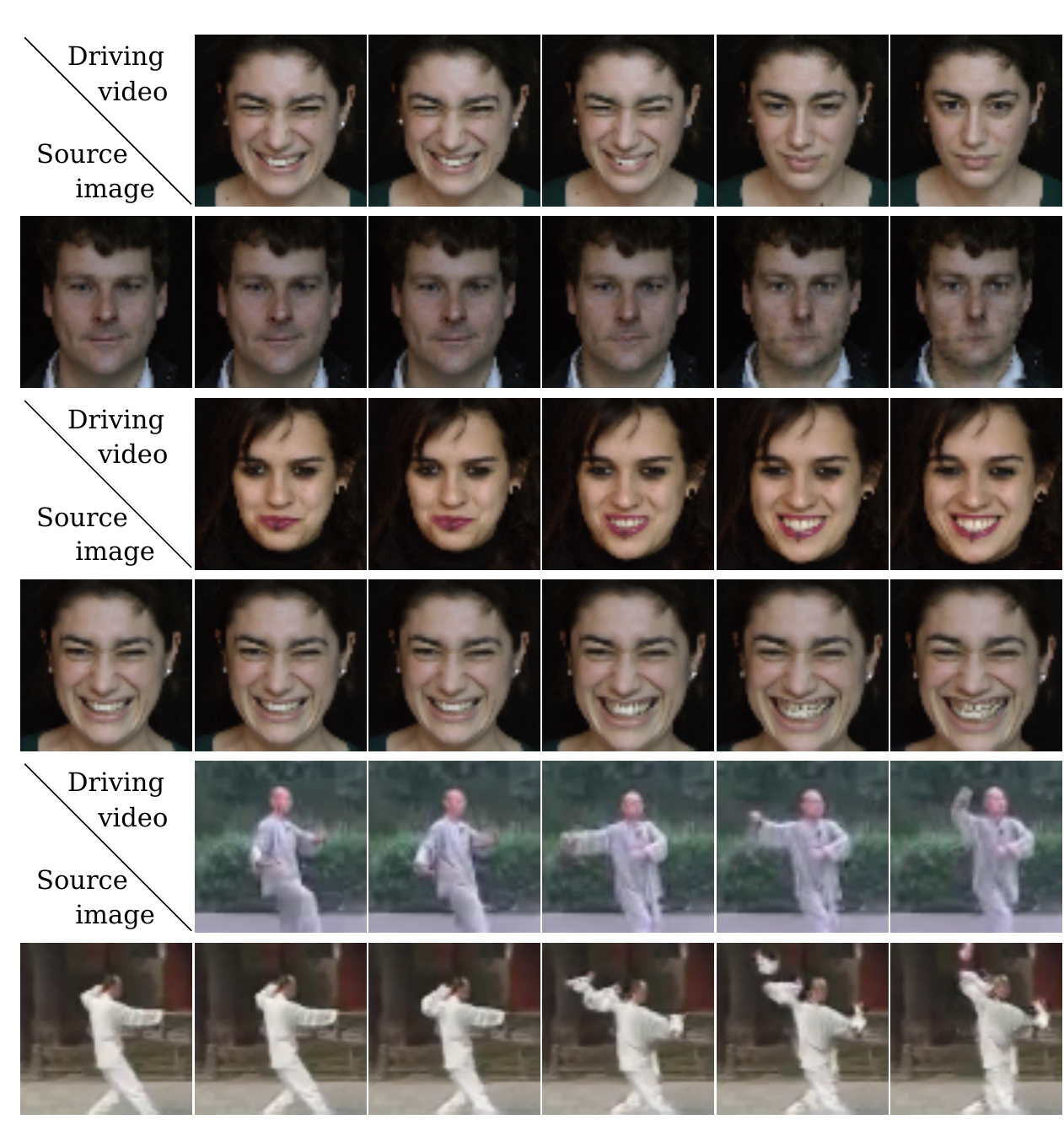}
  \caption{Illustration of the pose misalignment issue on the \emph{Nemo} and \emph{Tai-Chi} datasets.}
\label{fig:fail}
\end{figure}

Our approach assumes that the object in the first frame of the driving video and the object in the source image should be in similar poses. This assumption was made to avoid situations of meaningless motion transfer as shown in Fig.~\ref{fig:fail}. In the first row, the driving video shows the action of closing the mouth. Since the mouth of the subject in the source image is already closed, mouth disappears in the generated video. Similarly, in the second row, the motion in the driving video shows a mouth opening sequence. Since the mouth is already open in the source image, motion transfer leads to unnaturally large teeth. In the third row the man is asked to raise a hand, while it has already been raised.

\subsection{Additional ablation study }
\label{sec:abla}

\begin{figure}[h]
  \centering
\includegraphics[width=0.8\columnwidth]{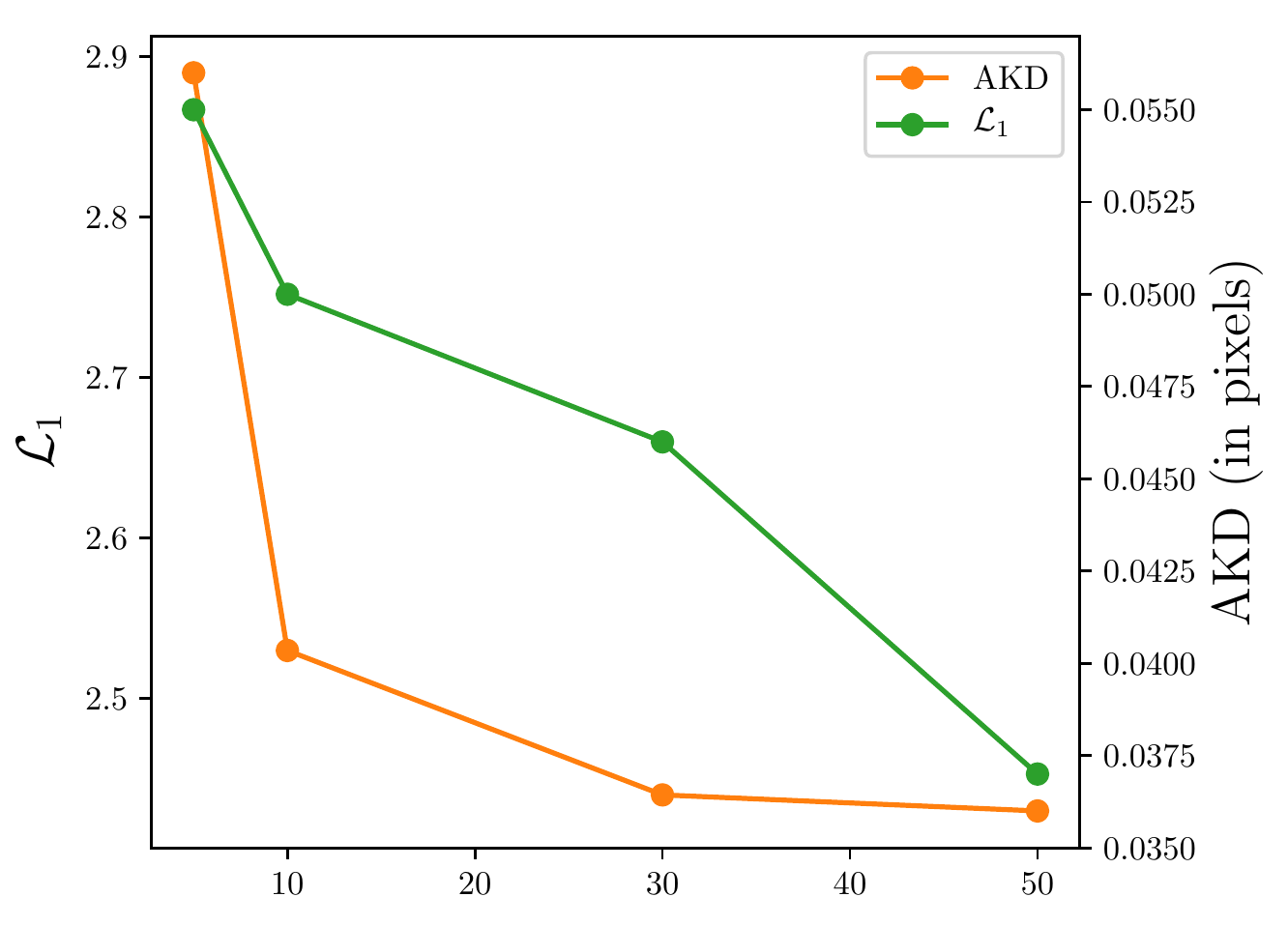}
    \caption{Reconstruction Errors as functions of the number of keypoints. Results obtained on \emph{Tai-Chi}.}
\label{fig:kp-curve}
\end{figure}

We perform experiments to measure the impact of the number of keypoints on video reconstruction quality. We report results on \emph{Tai-Chi} dataset in Fig.~\ref{fig:kp-curve}. We computed $\mathcal{L}_1$ and \emph{AKD} metrics as described in the paper. As expected, increasing the number of keypoints leads to a lower reconstruction error, but additional keypoints introduce memory and computational overhead. We use 10 keypoints in all our experiments, since we consider this to be a good trade-off.

\subsection{Image-to-Video translation}
\label{sec:i2v}

As explained in Sec.~4.2 of the main paper, we compare with the three state of the art methods for Image-to-Video translation: MoCoGAN~\cite{tulyakov2017mocogan} and SV2P~\cite{babaeizadeh2017stochastic}, and CMM-Net~\cite{wang2018every}. CMM-Net is evaluated only on \emph{Nemo} and SV2P only on the \emph{Bair} dataset. We report a user study and qualitative results. 

\noindent
\textbf{User Study.}
We perform a user study for the image-to-video translation problem. As explained in Sec.~4.3, we perform pairwise comparisons between our method and the competing methods. We employ the following protocol: we randomly select 50 videos and use the first frame of each of video as the reference frames to generate new videos. For each of the 50 videos the initial frame, and two videos generated by our and one of the competing methods are shown to the user. We provide the following instructions: "Select a more realistic animation of the reference image". As in Sec.~4.2 of the main paper, our method is compared with MoCoGAN \cite{tulyakov2017mocogan}, Sv2p~\cite{babaeizadeh2017stochastic}, and CMM-Net~\cite{wang2018every}.
The results of the user study are presented in Table~\ref{tab:prediction-user}. On average, users preferred the videos generated by our approach over those generated by other methods. The preference gap is especially evident for the \emph{Tai-Chi} and \emph{Bair} datasets that contain a higher amount of large motion. This supports the ability of our approach to handle driving videos with large motion.

\begin{table}[h!]
    \centering
    \begin{tabular}{c|c|c|c}
     \toprule
        &\emph{Tai-Chi}&\emph{Nemo}&\emph{Bair} \\
       
    \midrule    
        MoCoGAN \cite{tulyakov2017mocogan} & 88.2\% & 68.2\% & 90.6\%   \\
        CMM-Net \cite{wang2018every}& -& 63.6\% &- \\
        SV2P \cite{babaeizadeh2017stochastic}  &-&-& 98.8\%  \\
        \bottomrule
    \end{tabular}
    \caption{User study results on image-to-video translation. Proportion of times our approach is preferred over the competitors methods}
    \label{tab:prediction-user}
\end{table}

\begin{figure}[h]
\vspace{-3cm}
  \centering
\includegraphics[width=1.05\columnwidth]{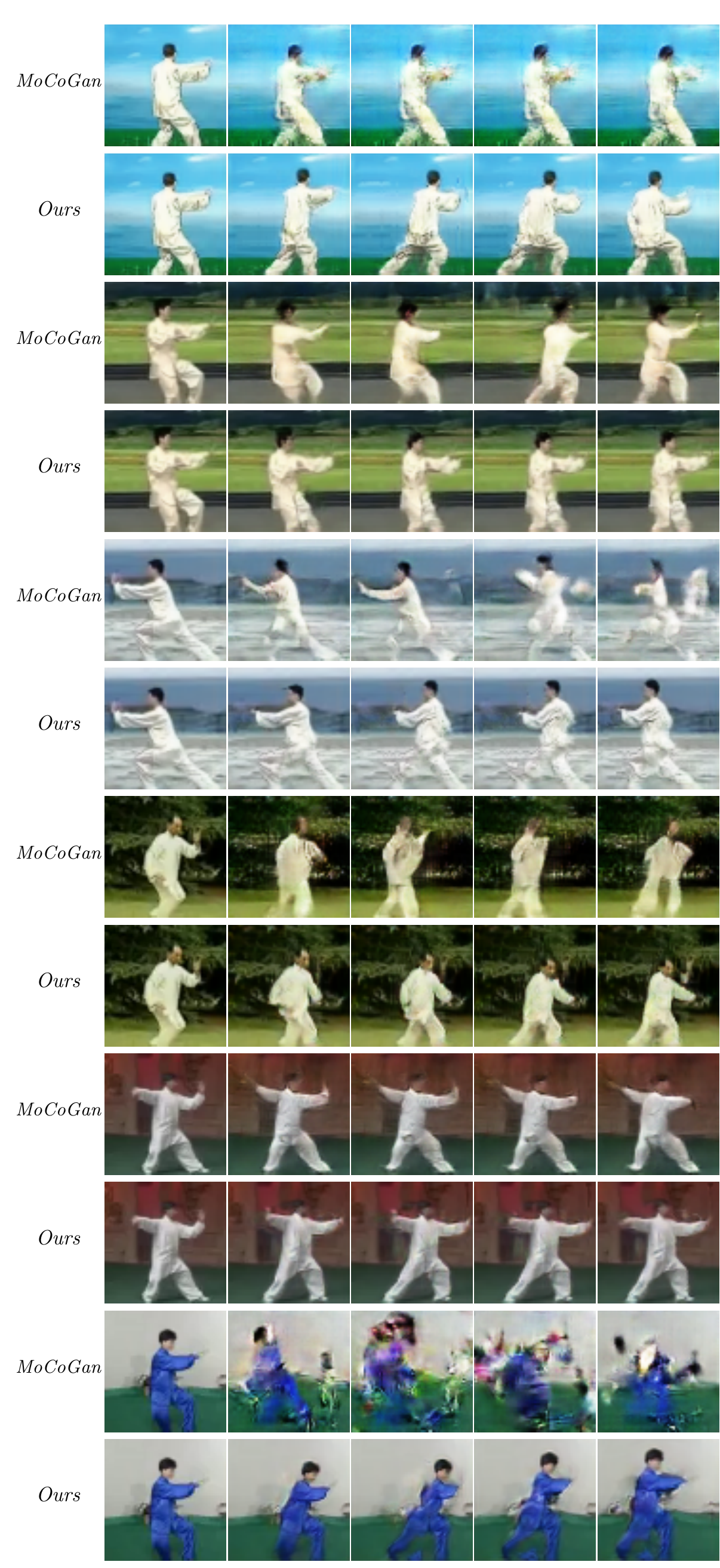}
  \caption{Qualitative results for \emph{Image-to-video}  translation on the \emph{Tai-Chi} dataset.}
\label{fig:taichi-prediction}
\end{figure}

\begin{figure*}[h]
  \centering
\includegraphics[width=2.1\columnwidth]{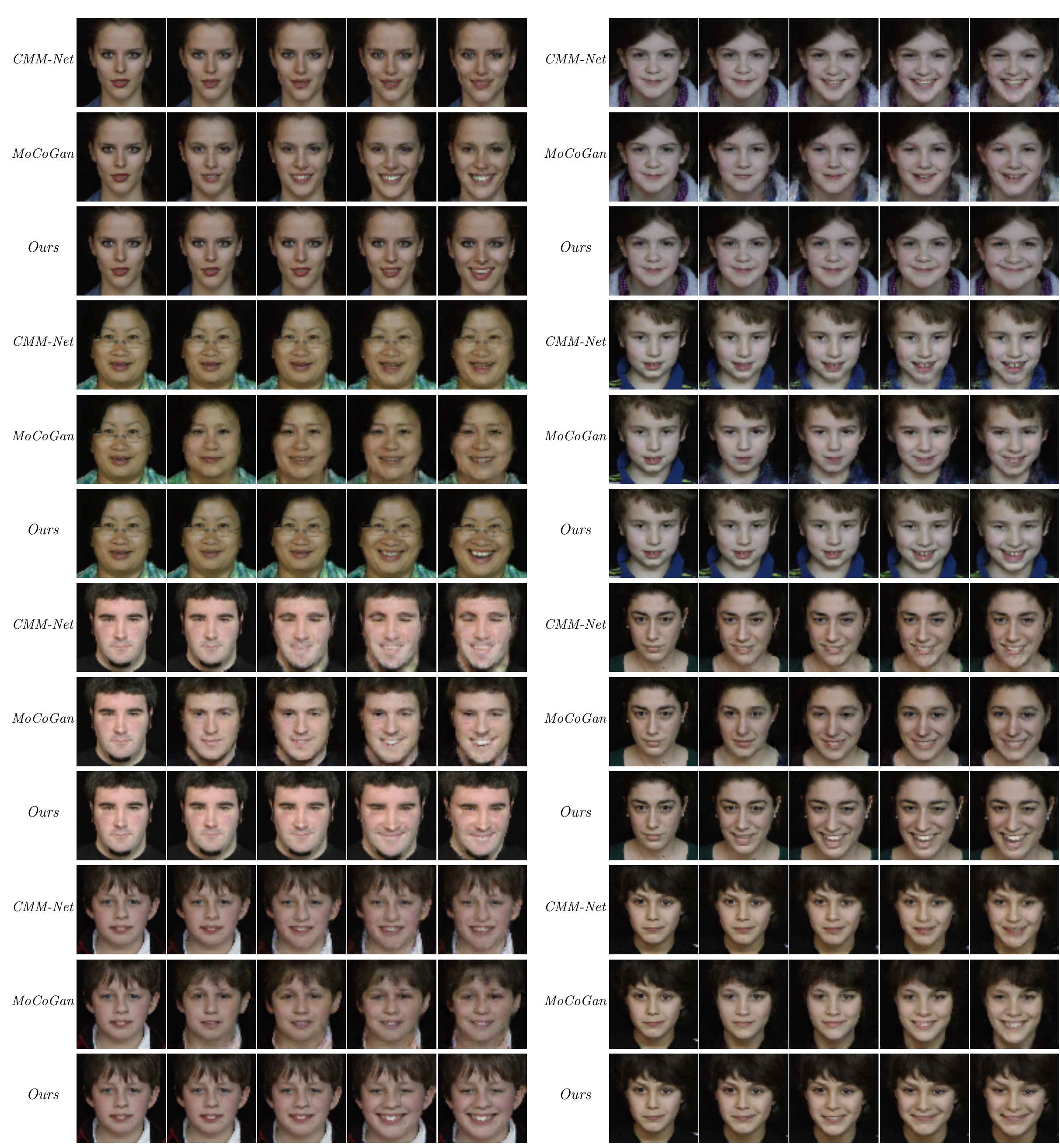}
    \caption{Qualitative results for \emph{Image-to-video}  translation on the \emph{Nemo} dataset.}
\label{fig:nemo-prediction}
\end{figure*}

\begin{figure*}[h]
  \centering
\includegraphics[width=2.1\columnwidth]{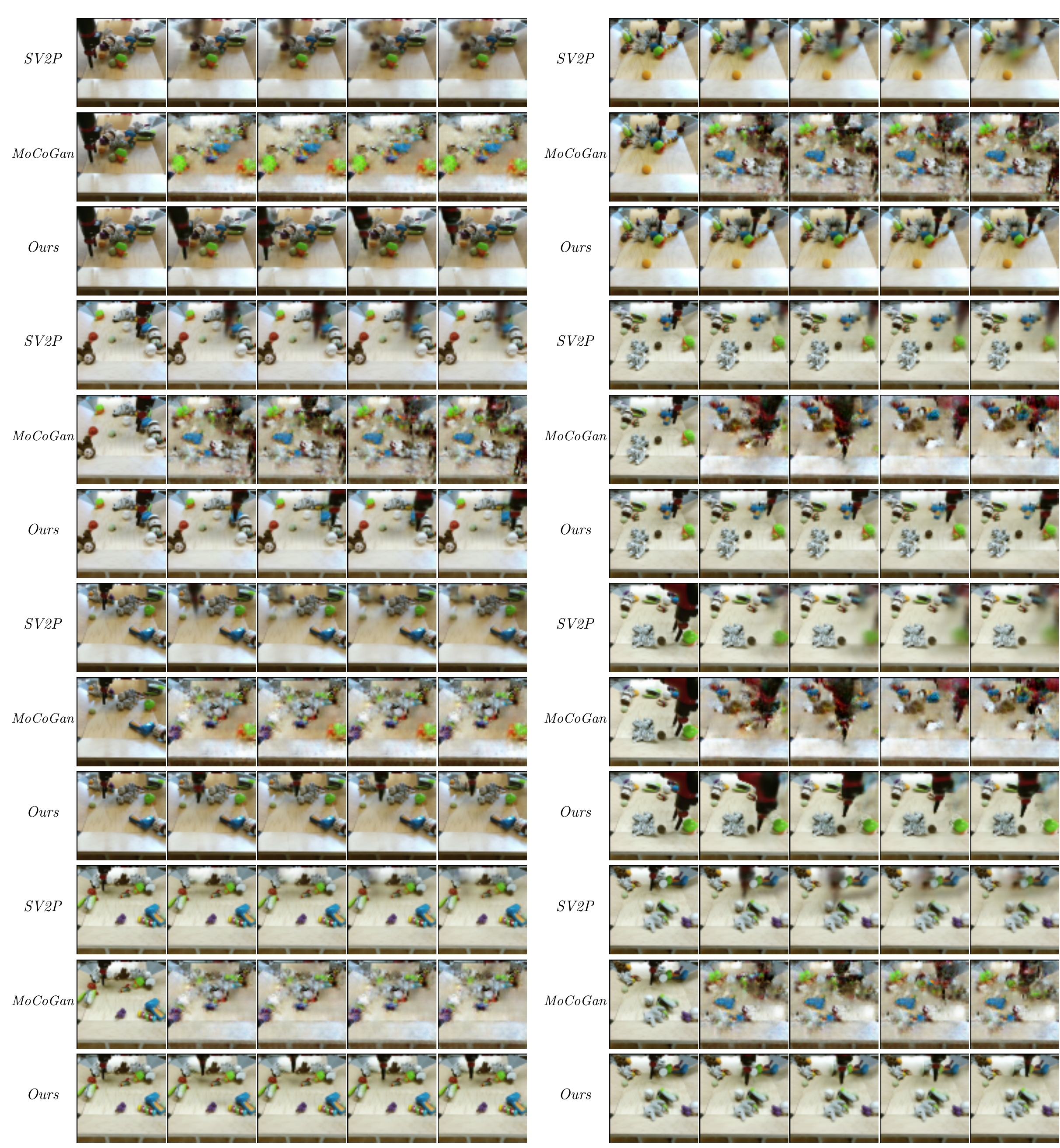}
    \caption{Qualitative results for \emph{Image-to-video} translation on the \emph{Bair} dataset.}
\label{fig:bair-prediction}
\end{figure*}

\noindent
\textbf{Qualitative results.}
We report additional qualitative results in Figs. \ref{fig:taichi-prediction}, \ref{fig:nemo-prediction}  and \ref{fig:bair-prediction}.
These qualitative results further support the ability of our method to generate realistic videos from source images and driving sequences. 

In particular, for the \emph{Nemo} dataset (Fig. \ref{fig:nemo-prediction}), MoCoGAN and CMM-Net suffer from more artifacts. In addition, the videos generated by MoCoGAN do not preserve the identity of the person. This issue is particularly visible when comparing the first and the last frames of the generated video. CMM-Net preserves better the identity but fails in generating realistic eyes and teeth. In contrast to these works, our method generates realistic smiles while preserving the person identity.

For \emph{Tai-Chi} (Fig. \ref{fig:taichi-prediction}), MoCoGAN~\cite{tulyakov2017mocogan} produces videos where some parts of the human body are not clearly visible (see rows 3,4 and 6). This is again due to the fact that visual information is embedded in a vector. Conversely, our method generates realistic human body with richer details.

For \emph{Bair} (Fig. \ref{fig:bair-prediction}), \cite{babaeizadeh2017stochastic} completely fails to produce videos where the robotic is sharp. The generated videos are blurred. MoCoGAN~\cite{tulyakov2017mocogan} generates videos with more details but containing many artifacts. In addition, the backgrounds generated by MoCoGAN are not temporally coherent. Our method generates realistic robotic arm moving in front of detailed and temporally coherent backgrounds.

\subsection{Image animation}
\label{sec:ia}
\begin{figure*}[h]
  \centering
\includegraphics[width=2\columnwidth]{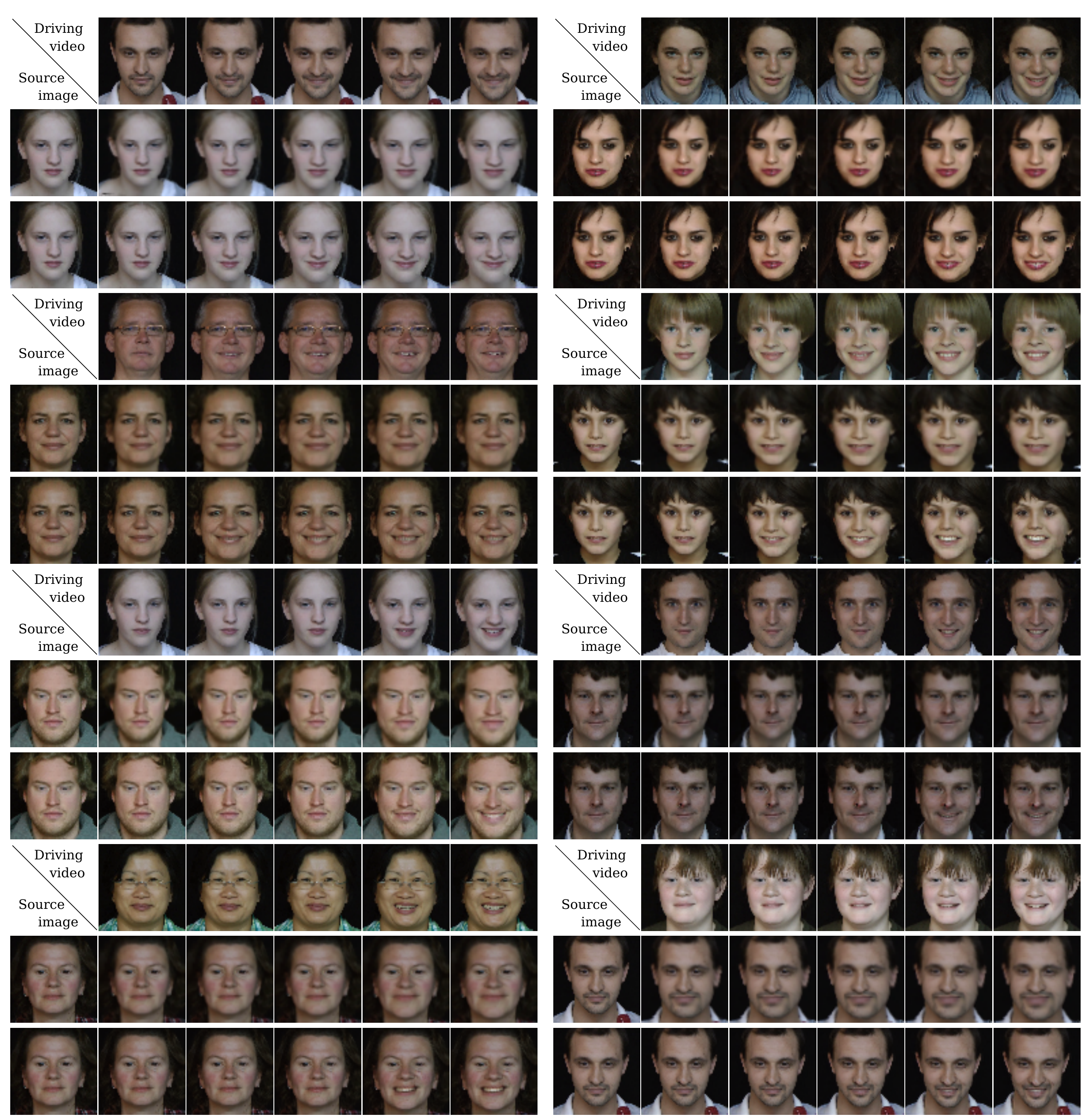}
  \caption{Additional qualitative results for image animation on the \emph{Nemo} dataset: X2face (first) against our method (second).}
\label{fig:nemo-transfer}
\end{figure*}

\begin{figure*}[h]
  \centering
\includegraphics[width=2\columnwidth]{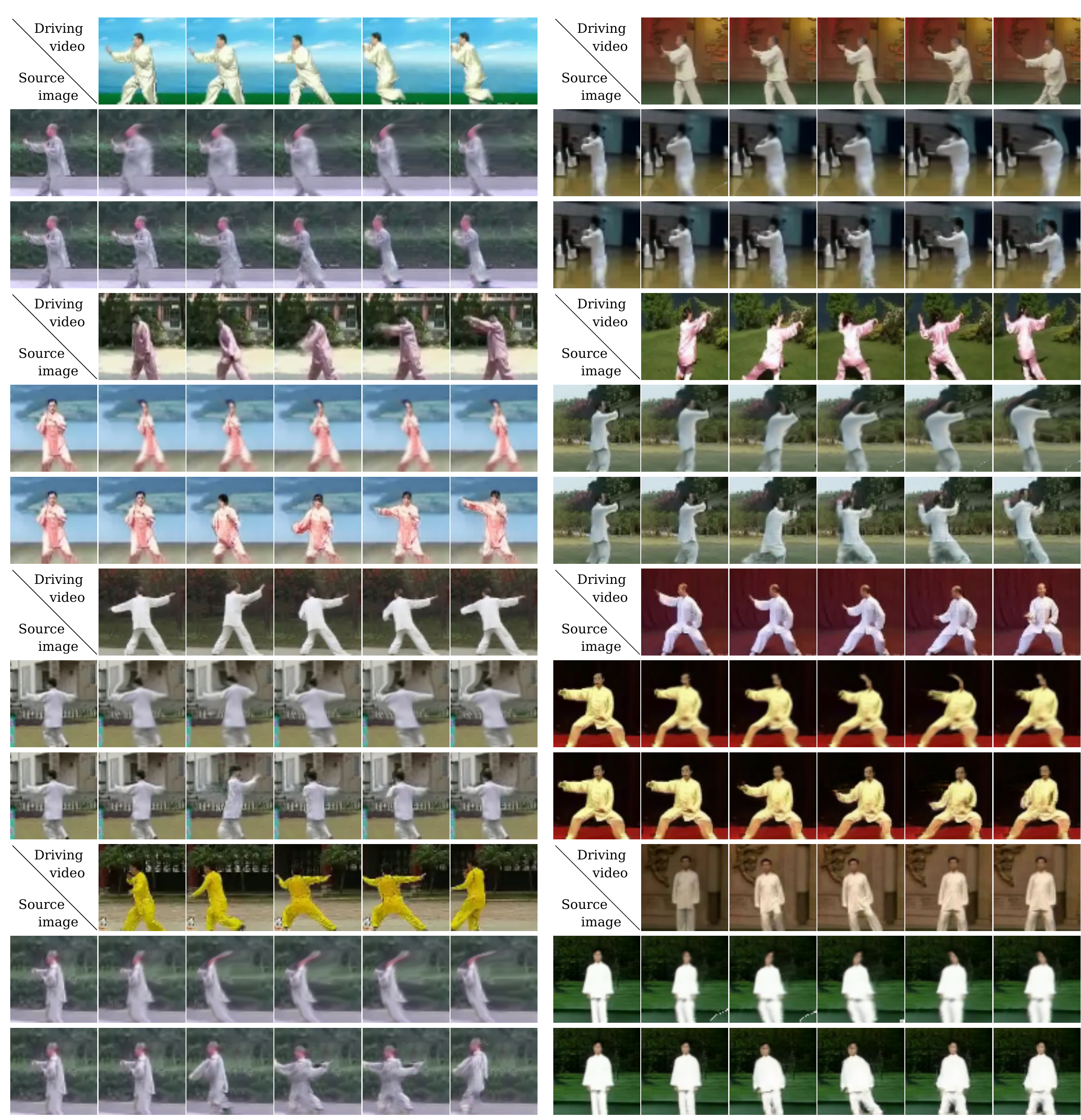}
  \caption{Additional qualitative results for image animation on the \emph{Tai-Chi} dataset: X2face (first) against our method (second).}
\label{fig:taichi-transfer}
\end{figure*}

\begin{figure*}[h]
  \centering
\includegraphics[width=2\columnwidth]{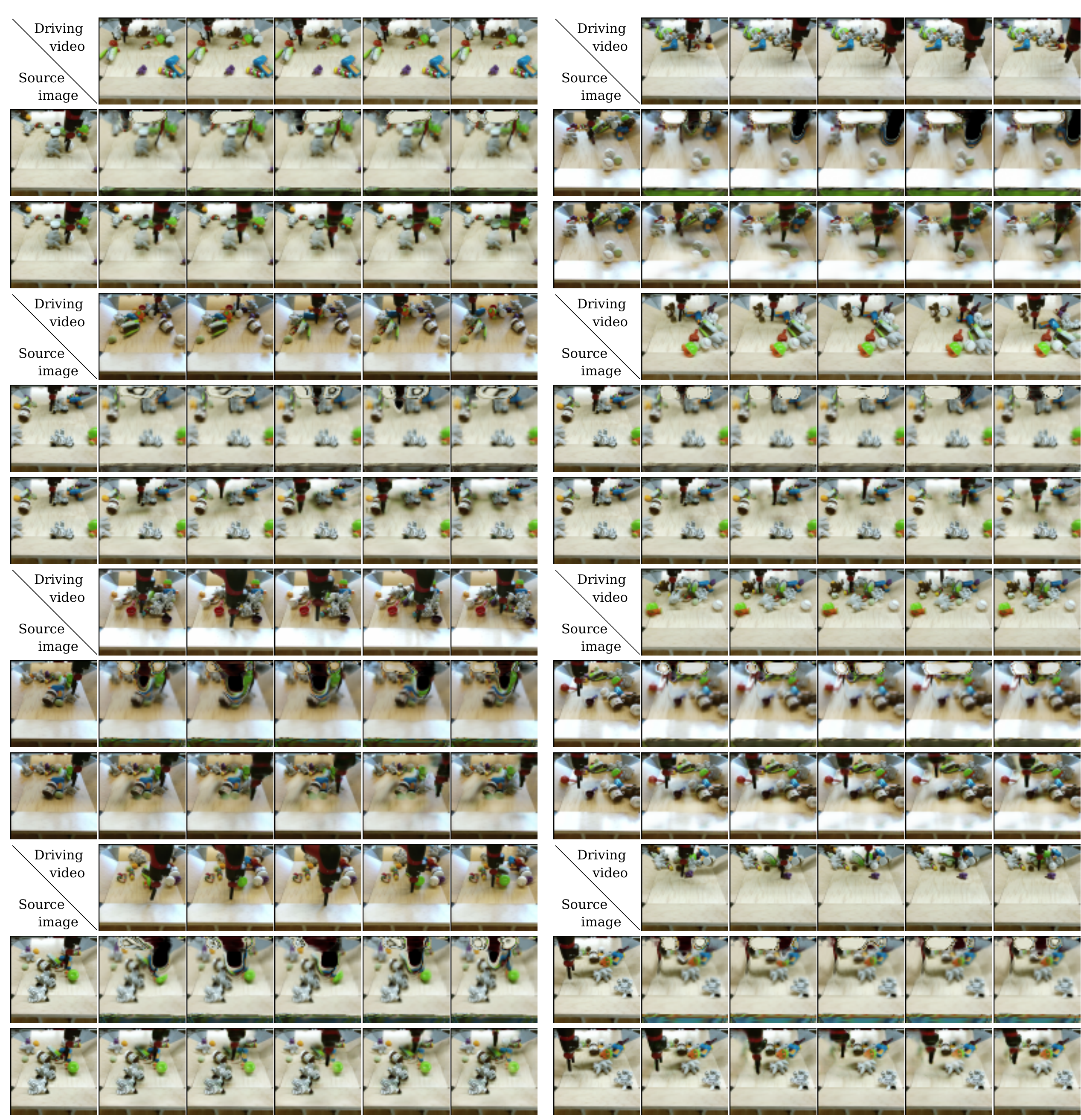}
  \caption{Additional qualitative results for image animation on the \emph{Bair} dataset: X2face (first) against our method (second).}
\label{fig:bair-transfer}
\end{figure*}

As explained in the main paper, we compare our method
with X2Face \cite{wiles2018x2face}. Results are reported in Figs.~\ref{fig:nemo-transfer},~\ref{fig:taichi-transfer} and ~\ref{fig:bair-transfer}  on the \emph{Nemo}, \emph{Tai-Chi} and \emph{Bair} datasets respectively.

When tested using the \emph{Nemo} dataset (Fig.~\ref{fig:nemo-transfer}), our method generates more realistic smiles on most of the randomly selected samples despite the fact that the XFace model is specifically designed for faces. Similarly to the main paper, the benefit of transferring the relative motion over absolute locations can be clearly observed in the bottom right example where the video generated by X2face inherits the large cheeks of the young boy in the driving video.

For \emph{Tai-Chi} (Fig.~\ref{fig:taichi-transfer}), X2face is not able to handle the motion of the driving video and simply warps the human body in the source image as a single blob.

For \emph{Bair} (Fig.~\ref{fig:bair-transfer}), we observe a similar behavior. X2face generates unrealistic videos where the robotic arm is generally not distinguishable. On the contrary, our model is able to generate a realistic robotic arm moving according to the driving video motion.

\begin{figure}[h]
  \centering
\includegraphics[width=1.05\columnwidth]{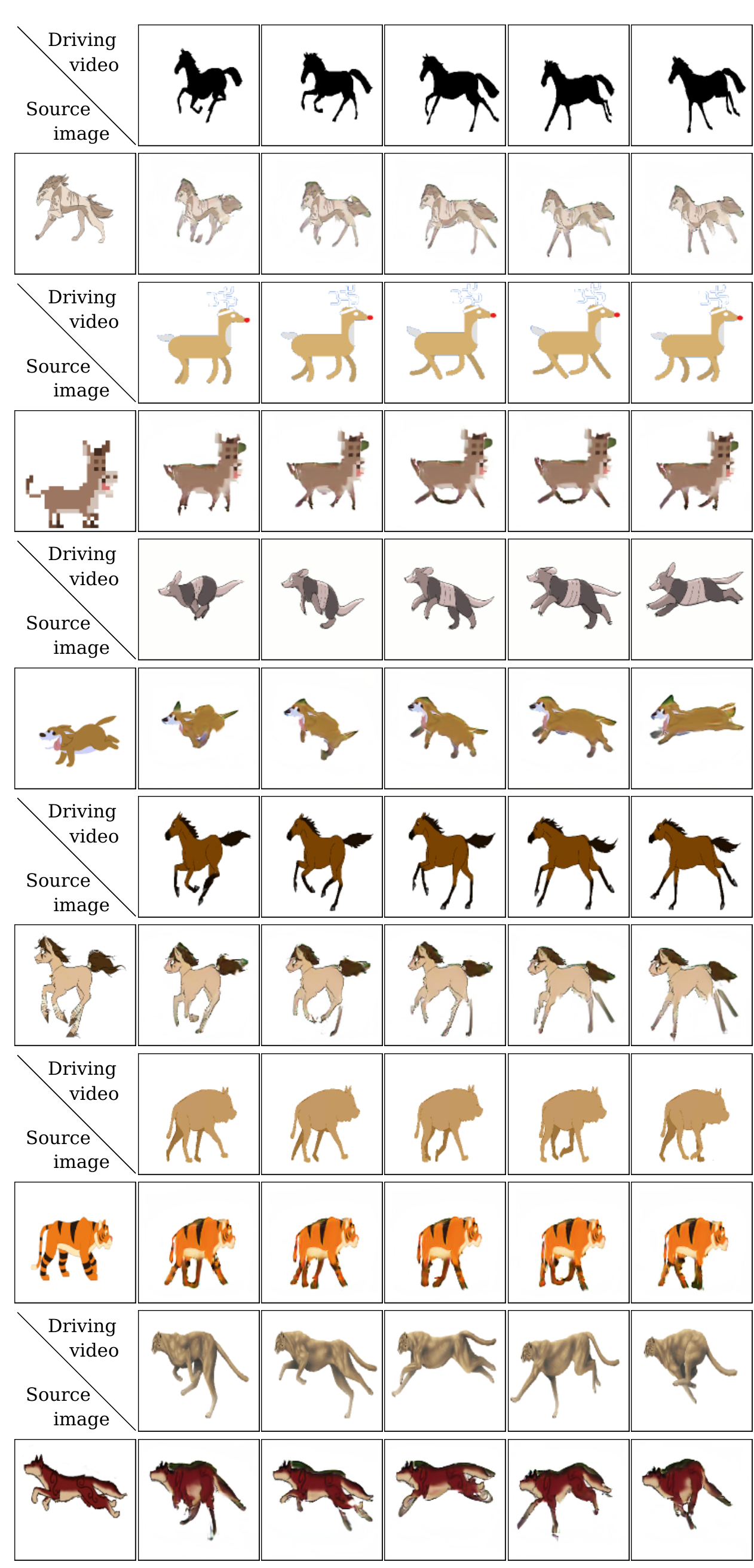}
  \caption{Qualitative results for image animation on the \emph{MGif}.}
\label{fig:moving-gif-transfer}
\end{figure}

Finally in Fig~\ref{fig:moving-gif-transfer}, we report results on the  \emph{MGif} dataset. First, these examples illustrate high diversity of \emph{MGif} dataset. Second, we observe that our model is able to transfer the motion of the driving video even if the appearance of the source frame is very different from the driving video. In particular, in all the generated sequences, we observe that the legs are correctly generated and follow the motion of the driving video. The model preserves the rigid parts of the animals as, for instance, the abdomen. In the last row, we see that the model is also able to animate the fox tail according to the  motion of the cheetah tail. 

\subsection{Keypoint visualization}
\label{sec:kv}
\begin{figure}[h]
  \centering
\includegraphics[width=1.05\columnwidth]{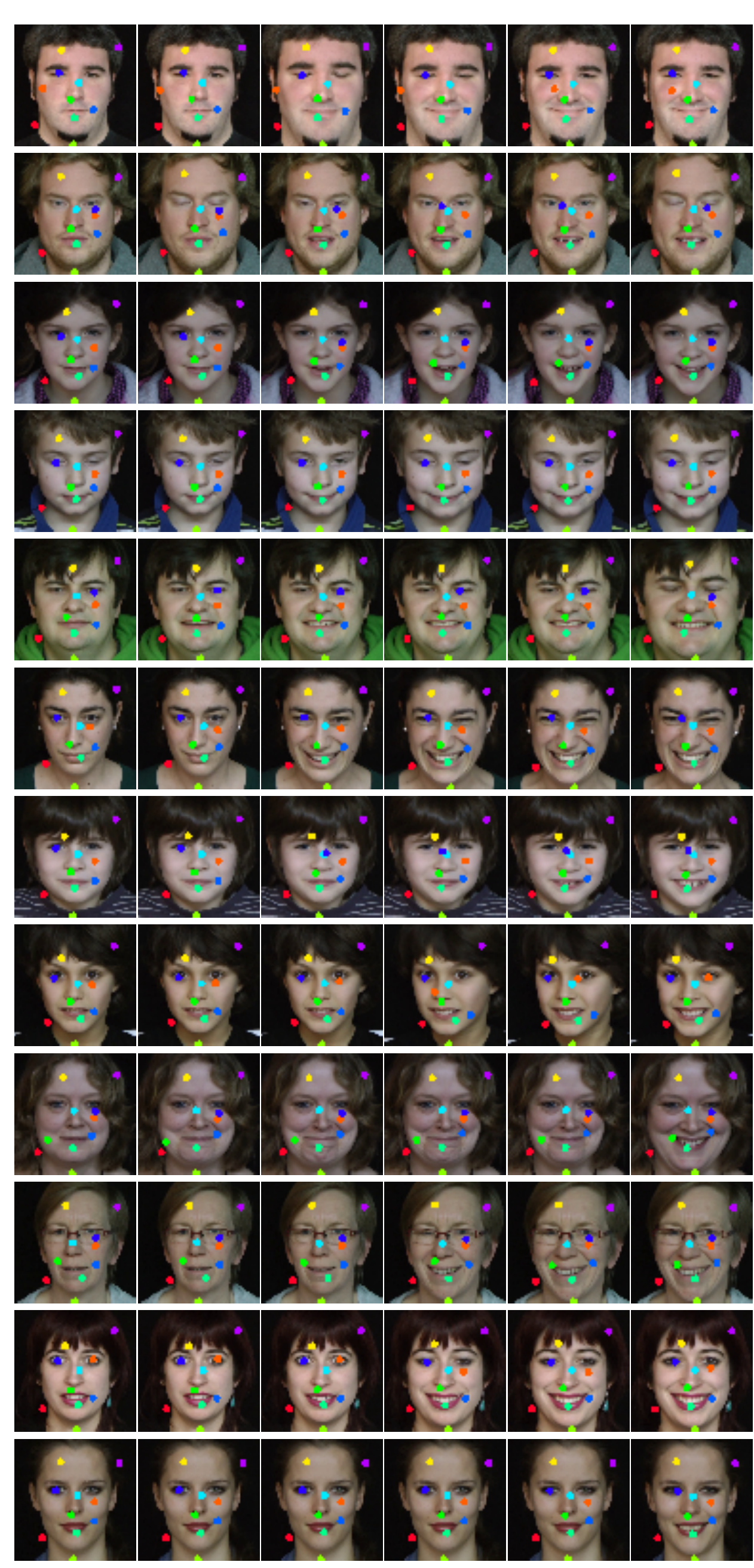}
  \caption{Keypoints predicted on the \emph{Nemo} dataset.}
\label{fig:knemo}
\end{figure}

\begin{figure}[h]
  \centering
\includegraphics[width=1.05\columnwidth]{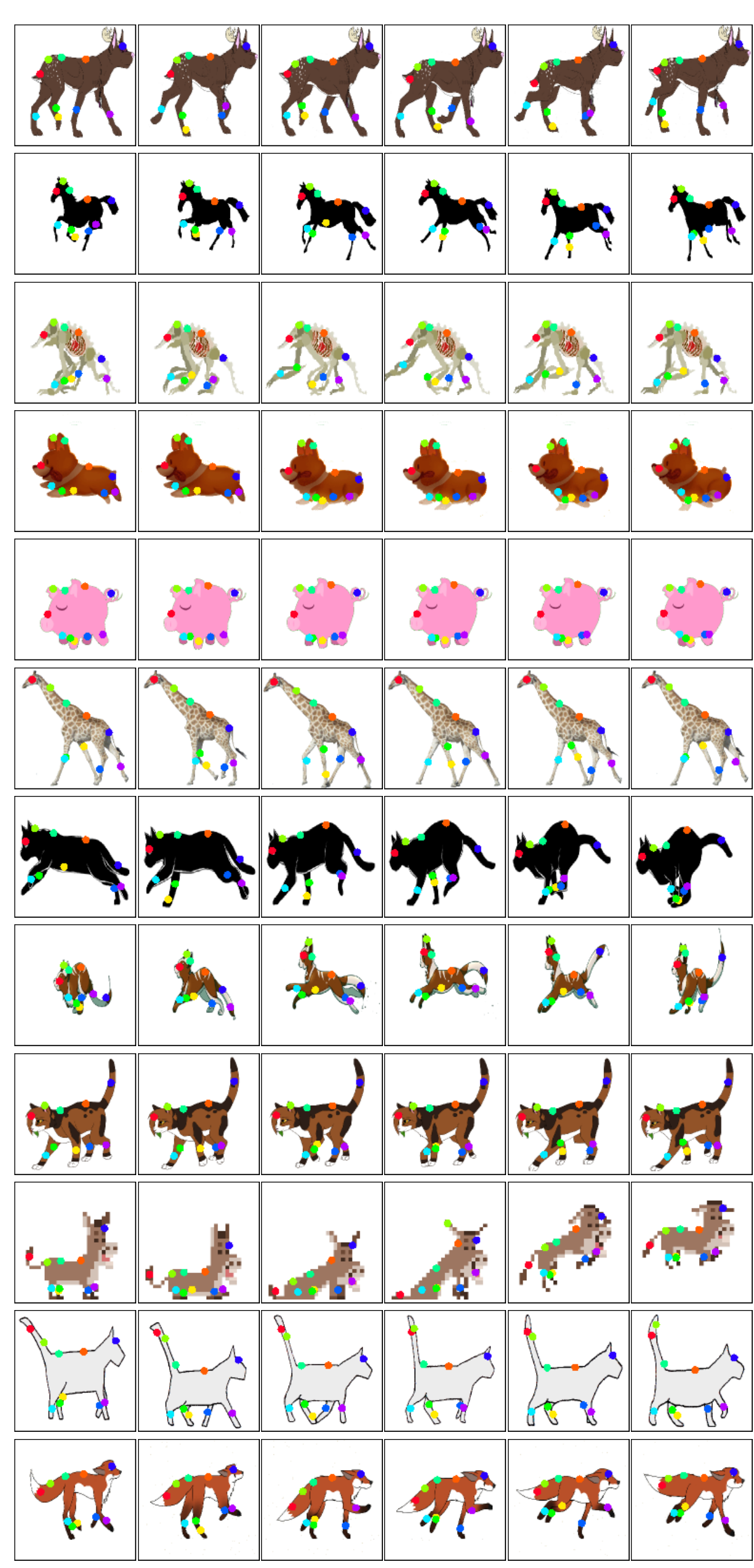}
  \caption{Keypoints predicted on the \emph{MGif} dataset.}
\label{fig:moving-gif-kp}
\end{figure}

\begin{figure}[h]
  \centering
\includegraphics[width=1.05\columnwidth]{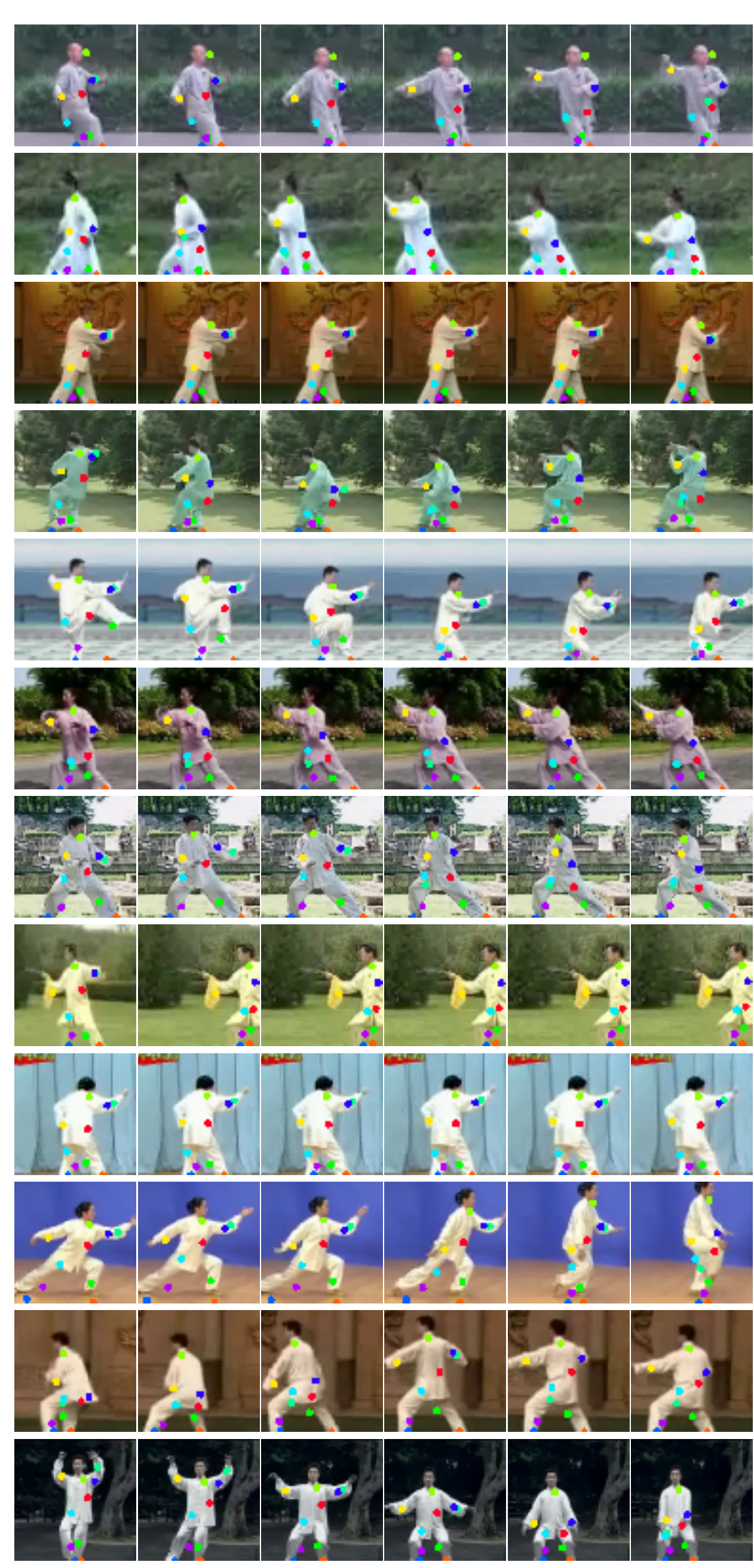}
  \caption{Keypoints predicted on the \emph{Tai-Chi} dataset.}
\label{fig:ktaichi}
\end{figure}

\begin{figure*}[h]
  \centering
\includegraphics[width=1.05\columnwidth]{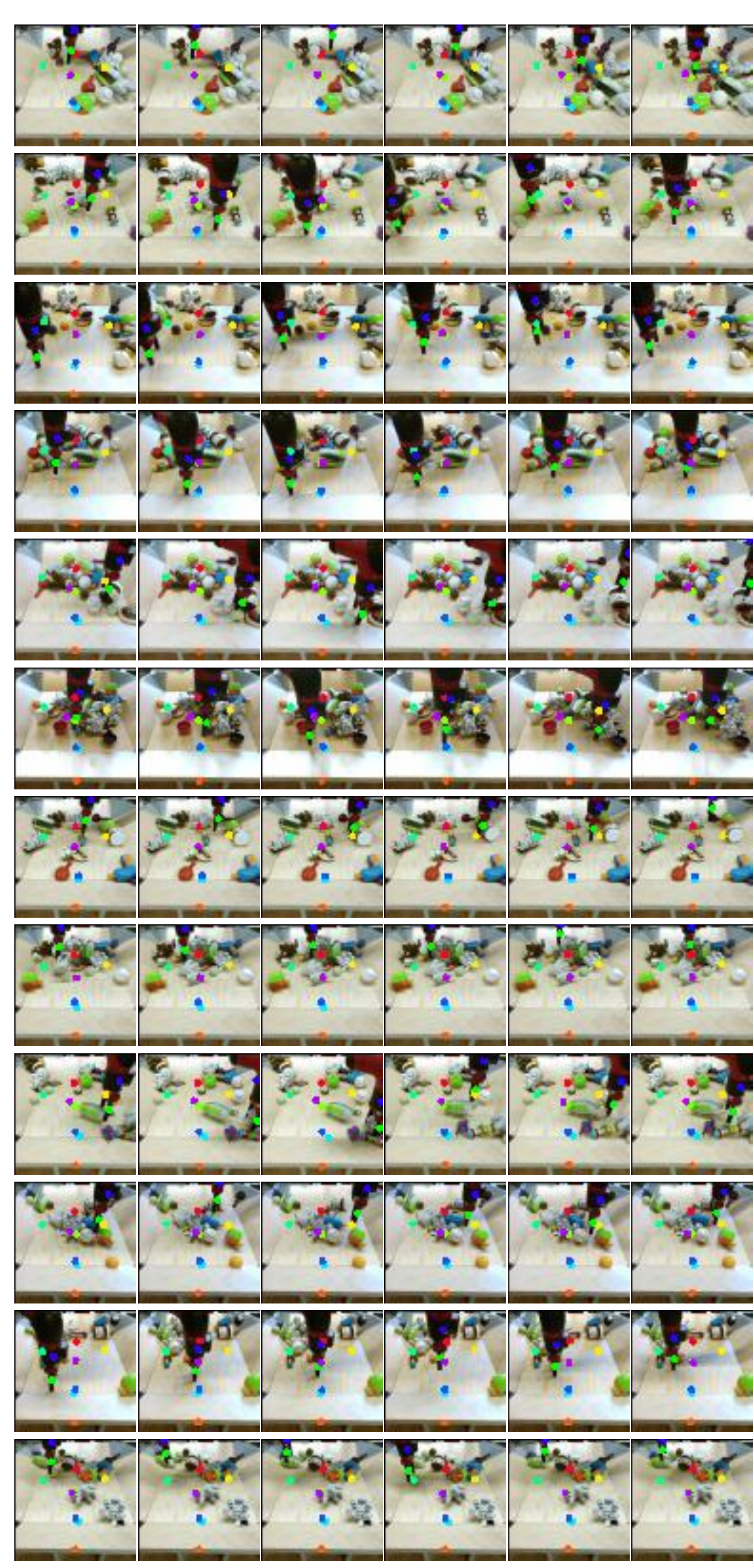}
  \caption{Keypoints predicted on the \emph{Bair} dataset.}
\label{fig:kbair}
\end{figure*}

 Finally, we report visual examples of keypoints learned by our model in Figs.~\ref{fig:knemo},~\ref{fig:moving-gif-kp},~\ref{fig:ktaichi} and \ref{fig:kbair}. On the \emph{Nemo} dataset, we observe that the obtained keypoints are semantically consistent. For instance, the cyan and light green keypoints constantly correspond to the nose and the chin respectively. For \emph{Tai-Chi}, the keypoints are also semantically consistent: light green for the chin and yellow for the left-side arm (right arm in frontal views and left arm in back views), for instance. For the \emph{Bair} dataset, we observe that two keypoints (light green and dark blue) correspond to the robotic arm. The other keypoints are static and can correspond to the background. Finally, concerning the \emph{MGif} dataset, we observe that each keypoint corresponds to two different animal parts depending if the animal is going towards left or right. In the case of animals going right (last three rows), the keypoints are semantically consistent (red for the tail, dark blue for the head etc.). Similarly, the keypoints are semantically consistent  among images of animal going left (red for the head, dark blue for the tail etc.). Importantly, we observe that a keypoint is associated to each highly moving part, as legs and tails.

\end{document}